\documentclass[aps,twocolumn,floats,prd,nofootinbib,superscriptaddress,10pt]{revtex4-1}

\usepackage{graphicx,amsmath,amssymb}
\usepackage{hyperref}
\hypersetup{colorlinks=true, linkcolor=red, filecolor=magenta, urlcolor=blue}

\AtBeginDocument{
    \newwrite\bibnotes
    \def\bibnotesext{Notes.bib}
    \immediate\openout\bibnotes=\jobname\bibnotesext
    \immediate\write\bibnotes{@CONTROL{REVTEX41Control}}
    \immediate\write\bibnotes{@CONTROL{
    apsrev41Control,author="08",editor="1",pages="1",title="0",year="1"}}
     \if@filesw
     \immediate\write\@auxout{\string\citation{apsrev41Control}}
    \fi
}

\begin{document}

\title{{\tt21cmFirstCLASS} II. Early linear fluctuations of the 21cm signal}

\author{Jordan Flitter}
\email{E-mail: jordanf@post.bgu.ac.il}
\author{Ely D.\ Kovetz}
\affiliation{Physics Department, Ben-Gurion University of the Negev, Beer-Sheva 84105, Israel}

\begin{abstract}
In a companion paper we introduce {\tt 21cmFirstCLASS}, a new code for computing the 21-cm anisotropies, assembled from the merger of the two popular codes {\tt 21cmFAST} and {\tt CLASS}. Unlike the standard {\tt 21cmFAST}, which begins at $z=35$ with homogeneous temperature and ionization boxes, our code begins its calculations from recombination, evolves the signal through the dark ages, and naturally yields an inhomogeneous box at $z=35$. In this paper, we validate the output of {\tt 21cmFirstCLASS} by developing a new theoretical framework which is simple and intuitive on the one hand, but is robust and precise on the other hand. As has been recently claimed, using consistent inhomogeneous initial conditions mitigates inaccuracies, which according to our analysis can otherwise reach the $\mathcal O\left(20\%\right)$ level.
On top of that, we also show for the first time that  {\tt 21cmFAST}  over-predicts the 21-cm power spectrum at $z\gtrsim20$ by another $\mathcal O\left(20\%\right)$, due to  the underlying assumption that $\delta_b=\delta_c$, namely that the density fluctuations in baryons and cold dark matter are indistinguishable. We propose an elegant solution to this discrepancy by introducing an appropriate scale-dependent growth factor into the evolution equations. Our analysis shows that this modification will ensure sub-percent differences between {\tt 21cmFirstCLASS} and the Boltzmann solver {\tt CAMB} at $z\leq50$ for all scales between the horizon and the Jeans scale. This will enable {\tt 21cmFirstCLASS} to consistently and reliably simulate  the 21-cm anisotropies both in the dark ages and cosmic dawn, for any cosmology. The code is publicly available at \href{https://github.com/jordanflitter/21cmFirstCLASS}{https://github.com/jordanflitter/21cmFirstCLASS}.
\end{abstract}

\maketitle

\section{Introduction}\label{sec: introduction}

It is truly remarkable how much we can learn about our Universe from hyperfine transitions in neutral hydrogen atoms that once permeated the intergalactic medium (IGM). Each said transition results in the emission of a $\sim21$-cm wavelength photon that is redshifted as it propagates to Earth. This is the cosmic 21-cm signal~\cite{Madau:1996cs, Barkana:2000fd, Bharadwaj:2004it, Furlanetto:2006jb, Pritchard:2011xb, Bera:2022vhw, Shaw:2022fre}. 

Studies of the 21-cm signal have shown that the rich astrophysical information contained in it can be exploited to learn about the formation of the first galaxies and stars~\cite{Barkana:2004vb, Furlanetto:2006tf, Park:2018ljd, Munoz:2019hjh, Park:2019aul, Greig:2019tcg, Thyagarajan:2020kua, Qin:2020xyh, Qin:2020pdx, Reis:2020hrw, Ma:2023oko}, probe the epoch of reionization~\cite{Lidz:2008ry, Wiersma:2012ek, Jensen:2013fha, Meerburg:2013dua, Greig:2015qca, Greig:2017jdj, Kolopanis:2019vbl, Greig:2020suk, Ghara:2020syx, Watkinson:2021ctc, Rahimi:2021wom, Qin:2021gkn, Saxena:2023tue}, and trace the thermal history of the IGM~\cite{Chen:2003gc, Mesinger:2013nua, Ghara:2019kir, Greig:2020hty, Ghara:2021fqo, Maity:2021mhr, HERA:2021noe, HERA:2022wmy, Lazare:2023jkg}. All of that potential lies within the high frequency bands ($\sim50-200\,\mathrm{MHz}$) of the 21-cm signal which can be measured by ground-based detectors. Worldwide efforts in pursuit of the signal are ongoing by numerous collaborations. Some of them aim at detecting the 21-cm global signal, like the Experiment to Detect the Global reionization Signature (EDGES)~\cite{Monsalve:2019baw},  Shaped Antenna measurement of the background RAdio Spectrum (SARAS)~\cite{2021arXiv210401756N}, the Radio Experiment for the Analysis of Cosmic Hydrogen (REACH)~\cite{deLeraAcedo:2022kiu}, Large-Aperture Experiment to Detect the Dark Ages (LEDA)~\cite{2018MNRAS.478.4193P} and Probing Radio Intensity at high-Z from Marion (PRIzM)~\cite{2019JAI.....850004P}, while others, such as Low Frequency Array (LOFAR)~\cite{Mertens:2020llj}, the Murchison Widefield Array (MWA)~\cite{Yoshiura:2021yfx}, the Giant Metrewave Radio Telescope (GMRT)~\cite{Pal:2020urw}, the Precision Array for Probing the Epoch of Reionization (PAPER)~\cite{Parsons:2013dwa}, the Hydrogen Epoch of Reionization Array (HERA)~\cite{HERA:2021bsv} and the Square Kilometre Array (SKA)~\cite{Braun:2015zta} focus on the spatial structure of the signal. All of these experiments are limited to the cosmological epochs of cosmic dawn and afterwards, at redshifts $z\lesssim35$, since lower frequencies are obscured by the Earth's ionosphere. In the next decades, radio receivers deployed on the far side of the moon or in lunar orbit~\cite{2009NewAR..53....1J, 2019arXiv191108649B, Furlanetto:2019jso, Silk:2020bsr, Burns:2021ndk, Burns:2021pkx, Goel:2022jgw, Shi:2022zdx, Bertone:2023ojo} will give us access to the poorly constrained epoch of the dark ages, enabling us to challenge the very foundations of the standard model of cosmology ($\Lambda$CDM)~\cite{Loeb:2003ya}.

Codes that can compute the 21-cm anisotropies both at the cosmic dawn and the dark ages, like {\tt 21cmFirstCLASS}, will therefore become invaluable. This code, introduced in a companion paper~\cite{Flitter:2023mjj} (Paper I), is based on two popular codes widely used in the literature, {\tt 21cmFAST}\footnote{\href{https://github.com/21cmfast/21cmFAST}{github.com/21cmfast/21cmFAST}}~\cite{Mesinger:2010ne, Munoz:2021psm} and {\tt CLASS}\footnote{\href{https://github.com/lesgourg/class_public}{github.com/lesgourg/class\_public}}~\cite{Blas:2011rf}. Unlike the standard {\tt 21cmFAST}, which begins its calculations at $z=35$ with homogeneous temperature and ionization boxes, {\tt 21cmFirstCLASS} is able to begin from recombination and evolve the boxes through the dark ages. Hence, temperature and ionization fluctuations are naturally developed, leading to an inhomogeneous box at $z=35$. According to Ref.~\cite{Munoz:2023kkg} (hereafter referred as JBM23), these early temperature fluctuations can have an important impact on the 21-cm power spectrum at low redshifts.

Currently, the only state-of-the-art code that can calculate these early fluctuations is {\tt CAMB}\footnote{\href{https://github.com/cmbant/CAMB}{github.com/cmbant/CAMB}}, following the work of Ref.~\cite{Lewis:2007kz} (hereafter LC07). Unlike {\tt 21cmFirstCLASS}, {\tt CAMB} works in a completely different way; it does not evolve a coeval box, but rather solves the linear coupled Boltzmann-Einstein equations to obtain the fluctuations in Fourier space. This is achievable because in $\Lambda$CDM, above $z=35$, the fluctuations are expected to be linear, i.e.\ small, and thus higher terms in perturbation theory can be neglected. Assuming that all early fluctuations are indeed linear during the dark ages, consistency requires {\tt 21cmFirstCLASS} to agree with {\tt CAMB}.

In this paper, we 
first quantify the effect of including  temperature and ionization fluctuations in {\tt 21cmFirstCLASS} as early as recombination, thereby leading to  inhomogeneous initial conditions at  the onset of cosmic dawn.  
We then proceed to validate the output of {\tt 21cmFirstCLASS} during the dark ages, by comparing its output with {\tt CAMB}. Although a direct comparison between the two codes is the most straightforward approach to do so, it can be affected by different physical effects which are not taken into account in both codes, as well as different approximations that are adopted. 

Instead, we take an analytical approach, and develop a theoretical framework to predict the output of these two codes. This framework is arguably more tractable than the one presented in LC07 and so allows to gather new insights on the nature of the linear fluctuations of the 21-cm signal at the relevant scales for 21-cm interferometery.

\begin{figure}
\includegraphics[width=0.9\columnwidth]{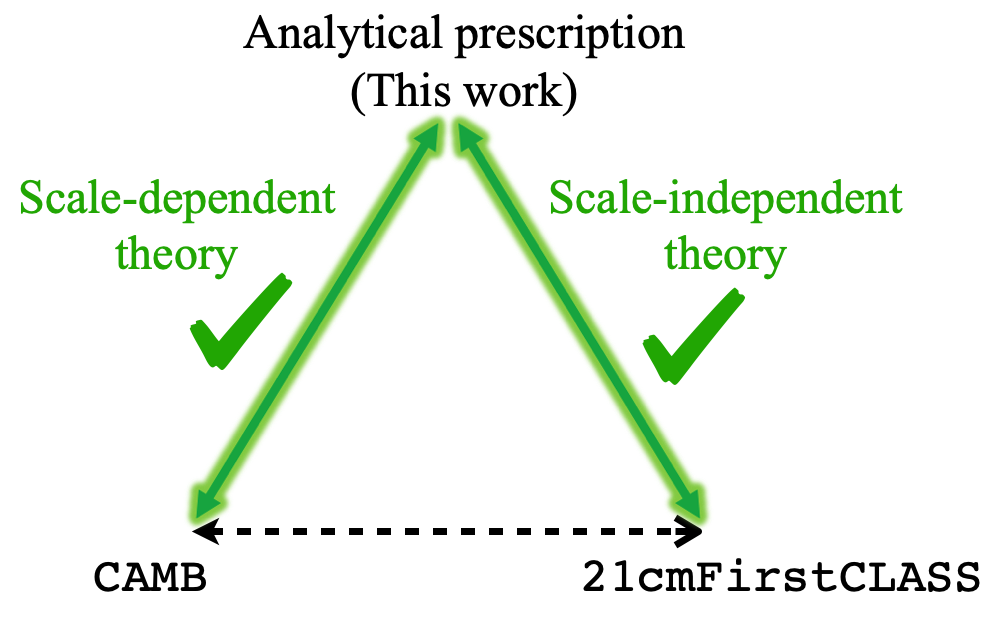}
\caption{Schematic representation of our comparison strategy. We compare between the analytical derivations of this work with {\tt 21cmFirstCLASS} ({\tt CAMB}), which uses a scale-independent (-dependent) theory for calculating the anisotropies in the 21-cm signal during the dark ages. As these comparisons are successful, our conclusion is that implementing the scale-dependent theory in {\tt 21cmFirstCLASS} will make it fully consistent with {\tt CAMB}.}
\label{Fig: figure_1}
\end{figure}

Our validation process is thus divided into three steps, as schematically presented in Fig.~\ref{Fig: figure_1}. First, we derive the linear fluctuations of all the relevant physical quantities during the dark ages. Secondly, we compare our scale-independent theory with the output of {\tt 21cmFirstCLASS}. And thirdly, we compare our scale-dependent theory with the output of {\tt CAMB}. Because the comparison in the last two steps is successful, we deduce that scale-dependent growth is the only ingredient missing in {\tt 21cmFirstCLASS} to make it consistent with {\tt CAMB}. This modification in the code of {\tt 21cmFirstCLASS} is feasible and will be implemented in its next version, which shall be made public.

The remaining parts of this paper are organized as follows. In Sec.~\ref{sec: 21cm Theory} we quote the standard equations used in 21-cm analysis. In Sec.~\ref{sec: Early temperature and ionization fluctuations} we analyze the evolution of early temperature and ionization fluctuations and their impact on the 21-cm power spectrum at low redshifts. In Sec.~\ref{sec: The 21cm signal during dark ages} we derive the brightness temperature fluctuations and relate them to the temperature and ionization fluctuations that were discussed at the previous section. In that section we also compare our scale-independent formalism with the output of {\tt 21cmFirstCLASS}. In Sec.~\ref{sec: non-linearities} we discuss the effect of non-linearities in the density field on the 21-cm power spectrum. We then compare our scale-dependent formalism with the output of {\tt CAMB} in Sec.~\ref{sec: Comparison with CAMB}. We conclude in Sec.~\ref{sec: Conclusions}.

As in Paper I, we adopt in this work the best-fit values for the cosmological parameters from Planck 2018 ~\cite{Planck:2018vyg} (without BAO), namely we assume a Hubble constant $h=0.6736$, total matter and baryons density parameters $\Omega_m=0.3153$, $\Omega_b=0.0493$, and a primordial curvature amplitude $A_s=2.1\times10^{-9}$ with a spectral index $n_s=0.9649$. For the fiducial values of the astrophysical parameters in {\tt 21cmFAST}, we adopt the EOS2021 values listed in Table 1 of Ref.~\cite{Munoz:2021psm}. To reduce clutter, we often do not explicitly write the independent arguments of the physical quantities (e.g.\ redshift, wavenumber, etc.) and they should be inferred from the context. All of our formulae are expressed in the CGS unit system.

\section{21cm Theory}\label{sec: 21cm Theory}

In Paper I we give a brief description of the 21-cm signal physics. Here we are only going to rewrite the fundamental equations that will serve us later in the derivations of this paper. Readers who wish to understand the origin of these equations can find classic reviews in Refs.~\cite{Madau:1996cs, Barkana:2000fd, Bharadwaj:2004it, Furlanetto:2006jb, Pritchard:2011xb}.

The 21-cm signal is the measurement of the physical quantity known as the brightness temperature,
\begin{equation}\label{eq: 1}
T_{21}=\frac{T_s-T_\gamma}{1+z}\left(1-\mathrm{e}^{-\tau_{21}}\right),
\end{equation}
where $T_\gamma\propto\left(1+z\right)$ is the redshift-dependent temperature of the cosmic microwave background (CMB), $T_s$ is the spin temperature, and $\tau_{21}\ll 1$ is the 21-cm optical depth,
\begin{equation}\label{eq: 2}
\tau_{21}=\frac{3\hbar A_{10}c\lambda_{21}^2\left(1-x_e\right)n_\mathrm{H}}{16Hk_BT_s}\left(1+\frac{1}{H}\frac{d\left(\mathbf{\hat n}\cdot\mathbf{v}_b\right)}{d\left(\mathbf{\hat n}\cdot\mathbf{x}\right)}\right)^{-1}.
\end{equation}
Here, $c$ is the speed of light, $\hbar$ is the (reduced) Planck constant, $k_B$ is the Boltzmann constant, $H$ is the redshift-dependent Hubble parameter, $\lambda_{21}\approx21\,\mathrm{cm}$ is the wavelength of a 21-cm photon, $A_{10}=2.85\times10^{-15}\,\mathrm{sec}^{-1}$ is the spontaneous emission coefficient from the excited hyperfine level to the ground state, and $x_e\equiv n_e/\left(n_\mathrm{H}+n_\mathrm{He}\right)$ is the ionization fraction where $n_e$, $n_\mathrm{H}$ and $n_\mathrm{He}$, are the free-electron, hydrogen-nuclei and helium-nuclei number-densities, respectively. Finally, the last term accounts for the comoving derivative of the baryons peculiar velocity\footnote{Note that in our notation, $\mathbf v_b\equiv d\mathbf x_b/dt$ is the comoving peculiar velocity of the baryons. The proper comoving velocity is obtained by multiplying that quantity with the scale factor.} along the line-of-sight $\mathbf{\hat n}$.

In thermal equilibrium the spin temperature obeys 
\begin{equation}\label{eq: 3}
T_s^{-1}=\frac{x_\mathrm{CMB} T_\gamma^{-1}+x_\mathrm{coll} T_k^{-1}+\tilde x_\alpha T_\alpha^{-1}}{x_\mathrm{CMB}+x_\mathrm{coll}+\tilde x_\alpha},
\end{equation}
where $T_k$ is the IGM gas kinetic temperature, $T_\alpha\approx T_k$ is the color temperature of Ly$\alpha$ photons, and $x_\mathrm{CMB}=\left(1-\mathrm{e}^{-\tau_{21}}\right)/\tau_{21}\sim1$, $\tilde x_\alpha$ and $x_\mathrm{coll}$ are the CMB~\cite{Venumadhav:2018uwn}, Ly$\alpha$~\cite{Hirata:2005mz}, and collisional~\cite{Furlanetto:2006jb} couplings, respectively. The evolution of $T_k$ is determined from
\begin{eqnarray}\label{eq: 4}
\nonumber\frac{dT_k}{dz}&=&\frac{dt}{dz}\Bigg[-2HT_k+\Gamma_C\left(T_\gamma-T_k\right)+\frac{2}{3}\frac{T_k}{1+\delta_b}\frac{d\delta_b}{dt}
\\&&\hspace{7mm}+\epsilon_\mathrm{ext}-\frac{T_k}{1+x_e}\frac{dx_e}{dt}\Bigg],
\end{eqnarray}
where $dt/dz=-\left[H\left(1+z\right)\right]^{-1}$, and $\Gamma_C$ is the Compton heating rate,
\begin{equation}\label{eq: 5}
\Gamma_C\equiv\frac{8\pi^2\sigma_T\left(k_BT_\gamma\right)^4}{45\hbar^3c^4m_e}\frac{x_e}{1+x_e},
\end{equation}
where $m_e$ is the electron mass and $\sigma_T$ is Thomson cross section. The term $\epsilon_\mathrm{ext}$ in Eq.~\eqref{eq: 4} denotes the heating rates from external sources, mainly X-ray heating (but Ly$\alpha$ and CMB heating rates~\cite{Reis:2021nqf, Venumadhav:2018uwn, Sarkar:2022dvl} can be included), and $\delta_b\equiv\delta\rho_b/\bar\rho_b$ is the contrast in the baryon-density fluctuations.

The last equation we need in order to study the fluctuations in the signal during the dark ages is the evolution equation for $x_e$. In general, this equation is very complicated and tracks the recombination states of both hydrogen and helium, while taking into account excitations to high-order energy levels. These effects have been implemented in the publicly available code {\tt HyRec}\footnote{\href{https://github.com/nanoomlee/HYREC-2}{github.com/nanoomlee/HYREC-2}}~\cite{Ali-Haimoud:2010hou, Lee:2020obi}, which we have incorporated in our {\tt 21cmFirstCLASS} code (see Paper I for more details). However, we will see in Sec.~\ref{sec: Early temperature and ionization fluctuations} that in order to derive analytically the evolution of temperature and ionization fluctuations, it is sufficient to consider the Peebles effective three-level atom model~\cite{1968ApJ...153....1P}, in which the evolution of $x_e$ is given by
\begin{equation}\label{eq: 6}
\frac{dx_e}{dz}=\frac{dt}{dz}\left[\left.\frac{dx_e}{dt}\right|_\mathrm{reio}+\mathcal C\left(\beta_\mathrm{ion}\left(1-x_e\right)-\alpha_\mathrm{rec}n_\mathrm{H}x_e^2\right)\right],
\end{equation}
where $\alpha_\mathrm{rec}$ is the recombination rate (in units of $\mathrm{cm^3/sec}$), $\beta_\mathrm{ion}$ is the early photoionization rate, and $\mathcal C$ is the Peebles coefficient (see Appendix~\ref{sec: Peebles fluctuations} for more details on the Peebles coefficient). The term $dx_e/dt|_\mathrm{reio}$ represents the reionization rate at late times. At early times (prior the epoch of reionization), the recombination rate and photoionization rates were in equilibrium, and thus
\begin{equation}\label{eq: 7}
\beta_\mathrm{ion}=\alpha_\mathrm{rec}\left(\frac{m_ek_BT_\gamma}{2\pi\hbar^2}\right)^{3/2}\mathrm{e}^{-\epsilon_0/\left(k_BT_\gamma\right)},
\end{equation}
where $\epsilon_0=13.6\,\mathrm{eV}$ denotes the ionization energy of the hydrogen atom from its ground state.

\section{Early temperature and ionization fluctuations}\label{sec: Early temperature and ionization fluctuations}

One of the advantages of {\tt 21cmFirstCLASS} is that it allows to study non-linear evolution above $z=35$ in models beyond $\Lambda$CDM. It is tempting however to examine if small, linear $\Lambda$CDM fluctuations prior to $z=35$ can induce a measurable impact on the brightness temperature during the cosmic dawn or afterwards.

To study the evolution of early linear temperature fluctuations, let us consider the temperature evolution equation, Eq.~\eqref{eq: 4}, prior to cosmic dawn, when $\epsilon_\mathrm{ext}=0$. We also neglect the last term in Eq.~\eqref{eq: 4}, and we will show that it has indeed anegligible effect on the analysis contained in this section. Thus 
\begin{equation}\label{eq: 8}
\frac{dT_k}{dz}=\frac{2T_k}{1+z}-\frac{\Gamma_C\left(T_\gamma-T_k\right)}{H\left(1+z\right)}+\frac{2}{3}\frac{T_k}{\left(1+\delta_b\right)}\frac{d\delta_b}{dz}.
\end{equation}
We now expand $T_k=\bar T_k+\delta T_k$, $\Gamma_C=\bar \Gamma_C+\delta \Gamma_C$ in linear perturbation theory. Throughout this work we shall assume that the CMB temperature is homogeneous, that is we assume $\delta T_\gamma\equiv0$. This is an excellent approximation because at the relevant subhorizon scales and redshifts the linear fluctuations in the CMB temperature are more negligible than all other fluctuations. The evolution equation for the background temperature $\bar T_k$ is similar to Eq.~\eqref{eq: 8}, though without the last term, while the evolution equation for $\delta T_k$ is
\begin{equation}\label{eq: 9}
\frac{d\delta T_k}{dz}=\frac{2\delta T_k}{1+z}+\frac{\bar\Gamma_C\delta T_k}{H\left(1+z\right)}-\frac{\delta\Gamma_C\left(T_\gamma-\bar T_k\right)}{H\left(1+z\right)}+\frac{2}{3}\bar T_k\frac{d\delta_b}{dz}.
\end{equation}
We see that the Compton heating term in Eq.~\eqref{eq: 8} induces the second and third terms on the RHS of Eq.~\eqref{eq: 9}, which we  refer to collectively as \emph{Compton fluctuations}. From Eq.~\eqref{eq: 5} it is straightforward to find that the linear fluctuations in the Compton heating rate are given by
\begin{equation}\label{eq: 10}
\delta\Gamma_C=\bar\Gamma_C\frac{\delta x_e}{\bar x_e\left(1+\bar x_e\right)},
\end{equation}
where $x_e=\bar x_e+\delta x_e$. Thus, Eq.~\eqref{eq: 9} becomes~\cite{Ali-Haimoud:2013hpa}
\begin{eqnarray}\label{eq: 11}
\nonumber\frac{d\delta T_k}{dz}&=&\frac{2\delta T_k}{1+z}+\frac{\bar\Gamma_C\delta T_k}{H\left(1+z\right)}-\frac{\bar\Gamma_C\left(T_\gamma-\bar T_k\right)}{H\left(1+z\right)}\frac{\delta x_e}{\bar x_e\left(1+\bar x_e\right)}
\\&&+\frac{2}{3}\bar T_k\frac{d\delta_b}{dz}.
\end{eqnarray}

For cold dark matter (CDM), the linear density fluctuations evolve in an almost scale-invariant fashion (especially at high redshifts before the baryons had settled in the CDM gravitational potential wells) according to $\delta_c\left(z\right)=D\left(z\right)\delta_0$, where $D\left(z\right)$ is the scale-\emph{independent} growth factor and $\delta_0\equiv\delta_c\left(z=0\right)$. For the baryons, we now 
define a scale-\emph{dependent} growth factor $\mathcal D_b\left(k,z\right)$ as
 \begin{equation}\mathcal D_b\left(k,z\right)\equiv\mathcal T_{b/c}\left(k,z\right)D\left(z\right),\end{equation} 
where $\mathcal T_{b/c}\left(k,z\right)\equiv\mathcal T_b\left(k,z\right)/\mathcal T_c\left(k,z\right)$ is the ratio between the baryon-density and the CDM-density transfer functions, so that  $\delta_b\left(k,z\right)=\mathcal T_{b/c}\left(k,z\right)\delta_c\left(z\right)=\mathcal D_b\left(k,z\right)\delta_0$.

\begin{figure}
\includegraphics[width=\columnwidth]{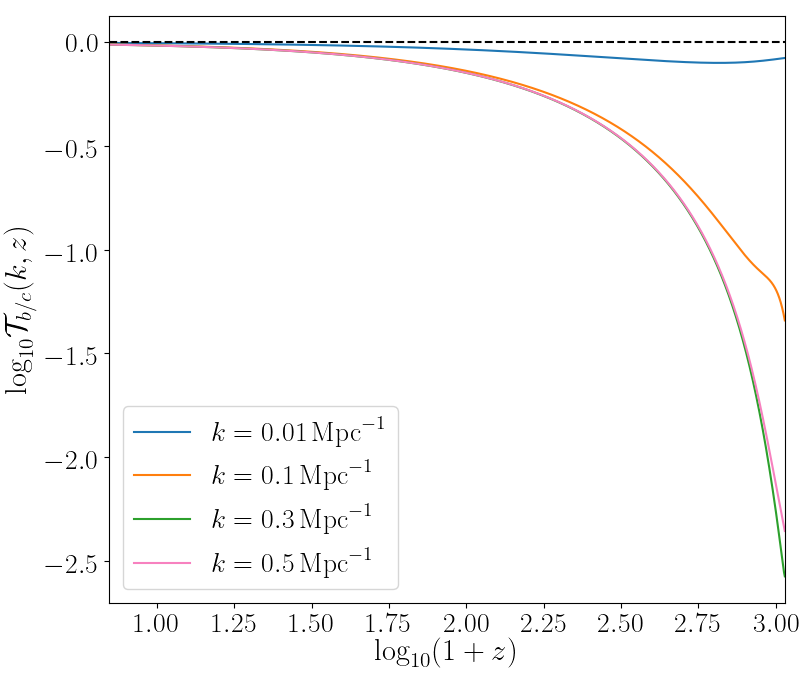}
\caption{The ratio between the baryons transfer function $\mathcal T_b(k,z)$ and the CDM transfer function $\mathcal T_c(k,z)$ as a function of redshift, for three different scales. For the smallest scales shown here (not including $k=0.01\,\mathrm{Mpc}^{-1}$), $\mathcal T_{b/c}(k,z)=0.97,\,0.96,\,0.92,\,0.88$ for $z=6,\,10,\,20,\,35$, respectively. At smaller scales, the resulting curve almost overlaps with the red curve of $k=0.5\,\mathrm{Mpc}^{-1}$. This figure was made with {\tt CLASS}.}
\label{Fig: figure_2}
\end{figure}

In Fig.~\ref{Fig: figure_2} we show $\mathcal T_{b/c}\left(k,z\right)$ as a function of redshift at the relevant scales for 21-cm interferometers. Initially, at high redshifts ($z\sim1000$), $\delta _b \ll \delta_c$, or $\mathcal T_{b/c}\left(k,z\right)\ll1$, since the baryons have recently been freed from the strong coupling with the CMB; before the baryons decoupled from the CMB photons, their density contrast was oscillating in a scale-dependent manner, while the CDM density contrast was continuously growing, especially after matter-radiation equality ($z\approx3300$) where $D\left(z\right)\propto\left(1+z\right)^{-1}$. Afterwards, for scales larger than the Jeans scale, the baryons can be considered as collisionless particles and they obey the same equation of motion as CDM. Therefore, for the scales of interest the baryons and the CDM become indistinguishable at low redshifts, after enough time has passed and their different initial conditions no longer play  an important role in their evolution. Hence, at low redshifts $\delta_b\to\delta_c$ and $\mathcal T_{b/c}\left(k,z\right)\to1$. This is the assumption made in the standard {\tt 21cmFAST}, where the simulation begins at $z=35$. 
However, the error induced by this assumption can be of the order of $\mathcal O\left(10\%\right)$ for $z\gtrsim20$. Since the 21-cm power spectrum $\Delta_{21}^2\left(k,z\right)$ is proportional to $\mathcal T_b^2\left(k,z\right)$, at least at redshifts prior to the non-linear cosmic dawn epoch (c.f. Eq.~\eqref{eq: 39}), this error becomes $\mathcal O\left(20\%\right)$ in the evaluation of $\Delta_{21}^2\left(k,z\right)$. Another interesting feature that can be seen in Fig.~\ref{Fig: figure_2} is that the relative suppression of $\mathcal T_b\left(k,z\right)$ at high redshifts is not necessarily stronger at smaller scales (or larger $k$). This is because different modes begin with different initial conditions, depending on their oscillatory phase at the moment of decoupling. However, adiabatic fluctuations imply $\delta_b\equiv\delta_c$ and $\mathcal T_{b/c}\left(k,z\right)\equiv1$ on superhorizon scales~\cite{Weinberg:2003sw}.

The solution for $\delta T_k$ in Eq.~\eqref{eq: 11} can be expressed analytically in the following way,
\begin{widetext}
\begin{eqnarray}\label{eq: 12}
\nonumber c_T\left(k,z\right)&=&-\frac{2}{3}\frac{\left(1+z\right)^2}{\bar T_k\left(z\right)\mathcal D_b\left(k,z\right)}\int_z^\infty dz'\frac{\bar T_k\left(z'\right)}{\left(1+z'\right)^2}\frac{d\mathcal D_b\left(k,z'\right)}{dz'}
\\\nonumber&&-\frac{\left(1+z\right)^2}{\bar T_k\left(z\right)\mathcal D_b\left(k,z\right)}\int_z^\infty dz'\frac{\bar \Gamma_C\left(z'\right)\bar T_k\left(z'\right)\mathcal D_b\left(k,z'\right)}{\left(1+z'\right)^3H\left(z'\right)}c_T\left(k,z'\right)
\\&&+\frac{\left(1+z\right)^2}{\bar T_k\left(z\right)\mathcal D_b\left(k,z\right)}\int_z^\infty dz'\frac{\bar \Gamma_C\left(z'\right)\left[T_\gamma\left(z'\right)-\bar T_k\left(z'\right)\right]\mathcal D_b\left(k,z'\right)}{\left(1+z'\right)^3H\left(z'\right)\left[1+\bar x_e\left(z'\right)\right]}c_{x_e}\left(k,z'\right),
\end{eqnarray}
\end{widetext}
where $c_T\equiv\delta T_k/\left(\bar T_k\cdot\delta_b\right)$ and $c_{x_e}\equiv\delta x_e/\left(\bar x_e\cdot\delta_b\right)$. Note that Eq.~\eqref{eq: 12} can be compared to Eq.~(47) in JBM23 where Compton fluctuations were neglected and the second and third terms are absent. Moreover, we see that $c_T$ is a scale-dependent quantity due to the non-trivial scale dependence of $\mathcal D_b\left(k,z\right)$. The first term in Eq.~\eqref{eq: 12} represents adiabatic temperature fluctuations due to structure growth. This term is positive because $d\mathcal D_b/dz<0$. The sign of the second and third terms is not clear at first sight and depends on the sign of $c_T$ and $c_{x_e}$. We claim however (and we verify this statement below) that these two terms give a \emph{negative} contribution to $c_T$ and thus lower its value. The second term is negative because $c_T>0$, as overdense regions are expected to be overheated. The third term is negative because $c_{x_e}<0$, as the recombination rate in overdense regions is higher and therefore the number of free electrons is reduced.

The fact that $c_T$ depends on the integrated histories of both $c_T$ and $c_{x_e}$ complicates its evaluation compared to the case where there are no Compton fluctuations. We will see below however that including all sources of fluctuations is absolutely necessary in order to obtain the correct solution.

In order to continue with the analytic derivation, we must include  the differential equation for $\delta x_e$. We shall not attempt to find the exact evolution equation for $x_e$, and instead we shall adopt the simpler Peebles model, Eq.~\eqref{eq: 6}. Omitting the late times reionization term, the evolution equation for $x_e$ reads
\begin{eqnarray}\label{eq: 13}
\nonumber\frac{dx_e}{dz}&=&\frac{\mathcal C}{H\left(1+z\right)}\left(\alpha_\mathrm{rec}n_\mathrm{H}x_e^2-\beta_\mathrm{ion}\left(1-x_e\right)\right)
\\&=&\left.\frac{dx_e}{dz}\right|_\mathrm{rec}+\left.\frac{dx_e}{dz}\right|_\mathrm{ion},
\end{eqnarray}
where we identified the recombination and ionization contributions to $dx_e/dz$. We then find, after working out the algebra, that~\cite{Ali-Haimoud:2013hpa}
\begin{eqnarray}\label{eq: 14}
\nonumber\frac{d\delta x_e}{dz}&=&\left(\frac{2}{\bar x_e}\left.\frac{d\bar x_e}{dz}\right|_\mathrm{rec}-\frac{1}{1-\bar x_e}\left.\frac{d\bar x_e}{dz}\right|_\mathrm{ion}\right)\delta x_e
\\&&+\left.\frac{d\bar x_e}{dz}\right|_\mathrm{rec}\delta_b+\left(\frac{\delta\mathcal C}{\overline{\mathcal C}}+\frac{\delta\alpha_\mathrm{rec}}{\bar\alpha_\mathrm{rec}}\right)\frac{d\bar x_e}{dz}.
\end{eqnarray}

In the name of brevity, we now ignore fluctuations in the Peebles coefficient $\mathcal C$, although they have non-negligible contributions to $\delta x_e$ at high redshifts. We provide more details on the calculation of $\delta\mathcal C$ in Appendix~\ref{sec: Peebles fluctuations}. In addition, we assume $\alpha_\mathrm{rec}$ is the case-B recombination rate and  a function of $T_k$ only, $\alpha_\mathrm{rec}=\alpha_B\left(T_k\right)$. With these assumptions, we can write Eqs.~\eqref{eq: 11} and \eqref{eq: 14} in matrix form,
\begin{flalign}\label{eq: 15}
\nonumber&\frac{d\delta v\left(k,z\right)}{dz}=A\left(z\right)\delta v\left(k,z\right)+B\left(k,z\right)&
\\&\hspace{15mm}=\left[\begin{matrix}
A_{11}\left(z\right) & A_{12}\left(z\right) \\
A_{21}\left(z\right) & A_{22}\left(z\right)
\end{matrix}
\right]\delta v\left(k,z\right)+\left[\begin{matrix}
B_{1}\left(k,z\right) \\
B_{2}\left(k,z\right)
\end{matrix}
\right],&
\end{flalign}
where $\delta v^\mathrm{T}=\left[\begin{matrix} \delta v_1 & \delta v_2\end{matrix}\right]\equiv\left[\begin{matrix} \delta T_k & \delta x_e\end{matrix}\right]$ and the elements of the matrices $A$ and $B$ are
\begin{eqnarray}
A_{11}\left(z\right)&=&\frac{2}{1+z}+\frac{\bar\Gamma_C}{H\left(1+z\right)}\label{eq: 16}
\\A_{12}\left(z\right)&=&-\frac{\bar\Gamma_C\left(T_\gamma-\bar T_k\right)}{H\left(1+z\right)}\frac{1}{\bar x_e\left(1+\bar x_e\right)}\label{eq: 17}
\\A_{21}\left(z\right)&=&\frac{\partial\ln\alpha_B\left(\bar T_k\right)}{\partial \bar T_k}\frac{d\bar x_e}{dz}\label{eq: 18}
\\A_{22}\left(z\right)&=&\frac{2}{\bar x_e}\left.\frac{d\bar x_e}{dz}\right|_\mathrm{rec}-\frac{1}{1-\bar x_e}\left.\frac{d\bar x_e}{dz}\right|_\mathrm{ion}\label{eq: 19}
\end{eqnarray}
\begin{eqnarray}
B_1\left(k,z\right)&=&\frac{2}{3}\bar T_k\frac{d\mathcal D_b\left(k,z\right)}{dz}\delta_0\label{eq: 20}
\\B_2\left(k,z\right)&=&\mathcal D_b\left(k,z\right)\left.\frac{d\bar x_e}{dz}\right|_\mathrm{rec}\delta_0,\label{eq: 21}
\end{eqnarray}
Then, Eq.~\eqref{eq: 15} has a closed form solution, which for zero initial conditions, $\delta v\left(k,z\to\infty\right)=0$, reads
\begin{equation}\label{eq: 22}
\delta v\left(k,z\right)=-\int_z^\infty dz'\exp\left[-\int_{z}^{z'}A\left(z''\right)dz''\right]B\left(k,z'\right).
\end{equation}
Note that in the absence of Compton fluctuations, Eq.~\eqref{eq: 22} is consistent with the first line of Eq.~\eqref{eq: 12} (and Eq.~(47) in JBM23). Because $B\left(k,z\right)\propto\delta_0$, $\delta v\left(k,z\right)\propto\delta_0$, and for the calculation of $c_T$ we can choose any arbitrary non-zero value of $\delta_0$ when we evaluate Eq.~\eqref{eq: 22}, we then get
\begin{eqnarray}
c_T\left(k,z\right)=\frac{\delta v_1\left(k,z\right)}{\bar T_k\left(z\right)\mathcal D_b\left(k,z\right)\delta_0}\label{eq: 23}
\\c_{x_e}\left(k,z\right)=\frac{\delta v_2\left(k,z\right)}{\bar x_e\left(z\right)\mathcal D_b\left(k,z\right)\delta_0}\label{eq: 24}.
\end{eqnarray}
Note that $c_{x_e}$ is an inevitable byproduct of the calculation of $c_T$.

In what follows, we solve numerically\footnote{While Eq.~\eqref{eq: 22} is analytically correct, we found it numerically challenging to evaluate the double integration to good precision.} Eq.~\eqref{eq: 15} from recombination, defined by $\bar x_e\left(z_\mathrm{rec}\right)\equiv0.1$, while assuming zero initial conditions~\cite{Ali-Haimoud:2013hpa}. In addition, in our analytical calculations we adopt the recombination model of {\tt RECFAST}~\cite{1991A&A...251..680P,Seager:1999bc,Seager:1999km},
\begin{equation}\label{eq: 25}
\alpha_B\left(T_k\right)=F_\alpha\frac{a_\alpha\left(T_k/10^4\,\mathrm{K}\right)^{b_\alpha}}{1+c_\alpha\left(T_k/10^4\,\mathrm{K}\right)^{d_\alpha}}\,\mathrm{\frac{cm^3}{sec}},
\end{equation}
where $a_\alpha=4.309\times10^{-13}$, $b_\alpha=-0.6166$, $c_\alpha=0.6703$, $d_\alpha=0.5300$, and  $F_\alpha=1.125$ is a fudge factor to reproduce the result of a multi-level atom calculation~\cite{, Lee:2020obi, 2010MNRAS.403..439R}.

\subsection{Scale independence}\label{subsec: Scale independence}

In order to gain some intuition for the qualitative features in the solution for $c_T$ and $c_{x_e}$, let us begin the discussion by assuming that these quantities evolve in a scale-independent manner, i.e.\ $c_T\left(k,z\right)=c_T\left(z\right)$, $c_{x_e}\left(k,z\right)=c_{x_e}\left(z\right)$. Mathematically speaking, this is equivalent to setting $\mathcal T_{b/c}\left(k,z\right)\equiv1$, valid only on superhorizon scales. However, it is important to bear in mind that on these scales the fluctuations in the CMB temperature cannot be ignored, as we do in our analysis. 

\begin{figure}[t!]
\includegraphics[width=\columnwidth]{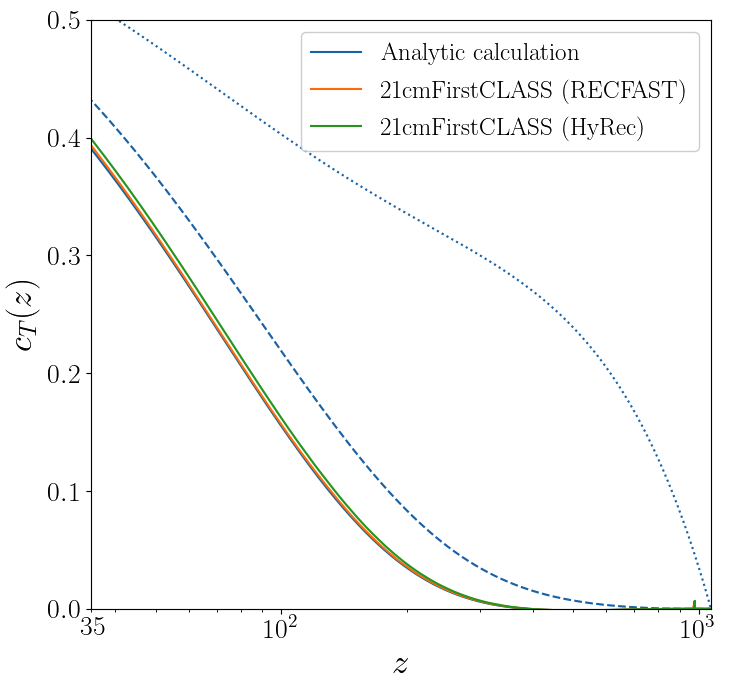}
\caption{The scale-\emph{independent} $c_T\left(z\right)$. The solid blue curve is the solution to Eq.~\eqref{eq: 15} (or Eq.~\eqref{eq: 22}) when accounting for Peebles fluctuations (see Appendix~\ref{sec: Peebles fluctuations}) while the orange (green) curve corresponds to the median of the $c_T\left(\mathbf x,z\right)$ box in {\tt 21cmFirstCLASS} when the considered recombination model is from {\tt RECFAST} ({\tt HyRec}). For comparison, we also show the contribution of Compton fluctuations to the solution. The blue dotted curve corresponds to accounting only for the first row in Eq.~\eqref{eq: 12}, the dashed blue curve corresponds to accounting for the first two rows in Eq.~\eqref{eq: 12}, and the solid blue curve corresponds to accounting for all three rows in Eq.~\eqref{eq: 12}.}
\label{Fig: figure_3}
\end{figure}

\begin{figure}[t!]
\includegraphics[width=\columnwidth]{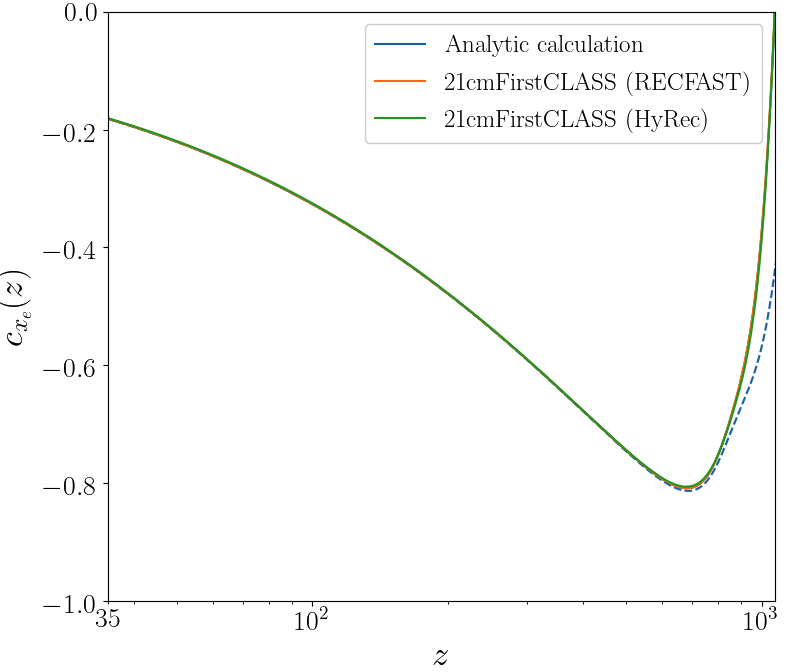}
\caption{The scale-\emph{independent} $c_{x_e}\left(z\right)$. The blue curve is the solution to Eq.~\eqref{eq: 15} (or Eq.~\eqref{eq: 22}) when accounting for Peebles fluctuations (see Appendix~\ref{sec: Peebles fluctuations}), while the orange (green) curve corresponds to the median of the $c_{x_e}\left(\mathbf x,z\right)$ box in {\tt 21cmFirstCLASS} when the considered recombination model is from {\tt RECFAST} ({\tt HyRec}). The calculation of all solid curves begins at $z_\mathrm{rec}=1069$ (corresponds to $\bar x_e\left(z_\mathrm{rec}\right)=0.1$), while the calculation of the blue dashed curve begins at $z=1200$.}
\label{Fig: figure_4}
\end{figure}

For the scale-independent evolution scenario, the solutions for $c_T\left(z\right)$ and $c_{x_e}\left(z\right)$ are presented by the solid blue curves in Figs.~\ref{Fig: figure_3} and \ref{Fig: figure_4}. As expected, $c_T\left(z\right)>0$ and $c_{x_e}\left(z\right)<0$. Therefore, as can be seen from Eq.~\eqref{eq: 12}, Compton fluctuations tend to decrease the value of $c_T\left(z\right)$ considerably, compared to the case when these fluctuations are not taken into account (represented by the blue dashed and dotted curves in Fig.~\ref{Fig: figure_3}). In addition, we see that Compton fluctuations are responsible for driving $c_T\left(z\right)$ to zero at high redshifts. This is physically well understood; at high redshifts, $z\gtrsim400$, $T_k$ is strongly coupled to the homogeneous $T_\gamma$ (see Fig.~1 in Paper I), hence $c_T\left(z\right)\approx0$ at these redshifts. And indeed, at $z\sim400$, once $T_k$ begins to depart from $T_\gamma$, $c_T\left(z\right)$ begins to grow via the source term of Eq.~\eqref{eq: 20}. Eventually, it reaches $\sim0.39$ at $z=35$. In fact, we find that at the vicinity of $z=35$, $c_T\left(z\right)$ can be approximated by the following fit,
\begin{equation}\label{eq: 26}
c_T\left(z\right)\approx0.39-0.0056\left(z-35\right)\qquad\text{(This work).}
\end{equation}
This fit is more precise than the fit found in JBM23 (originally derived in Ref.~\cite{Munoz:2015eqa})
\begin{equation}\label{eq: 27}
c_T\left(z\right)\approx0.43-0.006\left(z-35\right)\qquad\text{(JBM23).}
\end{equation}
The fit of JBM23, which is now used by default in the public version of {\tt 21cmFAST} (version 3.3.1) to account for early temperature fluctuations, is consistent with our calculations when the fluctuations in $x_e$ are ignored (the dashed curve in Fig.~\ref{Fig: figure_3} indeed reaches $0.43$ at $z=35$).

Unlike the monotonous growth of $c_T\left(z\right)$, the $c_{x_e}\left(z\right)$ curve exhibits a minimum at $z\sim700$. The origin for this minimum is due to the following. We find that $\delta_{x_e}\equiv\delta x_e/\bar x_e$ is monotonously decreasing towards lower negative values, but its rate of decrease is substantially higher above $z\sim700$. Since during matter domination the decrease in $D^{-1}\left(z\right)$ is roughly constant, $D^{-1}\left(z\right)\propto\left(1+z\right)$, which is faster (slower) than the decreasing rate of $\delta_{x_e}$ at $z\lesssim700$ ($z\gtrsim700$), the slope of $c_{x_e}\left(z\right)\propto\delta_{x_e}\left(z\right)D^{-1}\left(z\right)$ changes its sign at $z\sim700$. This does not happen in the $c_T\left(z\right)$ curve; between $35\leq z\leq 400$ the growth in $\delta_{T_k}\equiv\delta T_k/\bar T_k$ is always faster than the growth in $D\left(z\right)$.

Fig.~\ref{Fig: figure_4} displays an unphysical behavior; even though $c_{x_e}\left(z\right)$ begins at zero, its derivative does not. Of course, there is no reason to believe that $c_{x_e}\left(z\right)$ vanishes exactly at our initial redshift ($z=1069$), and it is an indication that the calculation has to begin at a higher redshift~\cite{Ali-Haimoud:2013hpa}. We show however in Fig.~\ref{Fig: figure_4} that if we begin the evolution at $z=1200$, the result remains the same after the turning point at $z\sim700$. Nevertheless, as we shall see in Sec.~\ref{sec: The 21cm signal during dark ages}, $c_{x_e}$ will only have a small  effect on the brightness temperature fluctuations at low redshifts, and we will therefore not concern ourselves with obtaining the correct solution for $c_{x_e}$ at redshifts higher than $z\sim700$.

We also make a comparison in Figs.~\ref{Fig: figure_3} and \ref{Fig: figure_4} between our analytical solution and the output from {\tt 21cmFirstCLASS}. To achieve the latter, we compute the redshift-dependent median of the coeval boxes $c_T\left(\mathbf x,z\right)$ and $c_{x_e}\left(\mathbf x,z\right)$ at each redshift iteration. When we disable {\tt HyRec} in {\tt 21cmFirstCLASS} and adopt our simple toy model for the evolution of $x_e$, the agreement between {\tt 21cmFirstCLASS} and the scale-independent theory is excellent, especially when inspecting the $c_{x_e}\left(z\right)$ curve. For $c_T\left(z\right)$, minor differences can be seen at low redshifts. There could be two reasons for this tiny discrepancy; either small non-linear fluctuations slightly shift the median of the $c_T\left(\mathbf x,z\right)$ box, or it could be the consequence of small numerical errors that increase as the box evolves. 

Regardless, these differences are smaller than the errors due to the approximated recombination model. Activating {\tt HyRec} in {\tt 21cmFirstCLASS} results in increasing slightly the value of $c_T\left(z\right)$, most likely due to a more precise evolution of $\bar T_k$ at high redshifts. Note however that because the absolute errors in $c_T\left(z\right)$ remain roughly constant over a wide redshift range, the relative errors are significantly larger at higher redshifts. Specifically, at $z\!=\!35$ ($z\!=\!100$) the error in $c_T\left(z\right)$ is $\sim\!1\%$ ($\sim4\%$). This small discrepancy will become relevant in our discussion in Sec.~\ref{sec: Comparison with CAMB}, where we make a comparison with {\tt CAMB}.

\subsection{The 21-cm power spectrum in {\tt 21cmFirstCLASS}}\label{subsec: The 21cm power spectrum in 21cmFirstCLASS}

Now that we have established the consistency of {\tt 21cmFirstCLASS}, at least when linear, scale-free, early temperature and ionizations fluctuations are considered, a natural question arises: do these early fluctuations  alter the observable, the 21-cm power spectrum at low redshifts, at a significant level? In other words, does the state of the box at $z=35$ matter for its evolution afterwards? To answer these questions, we calculate with {\tt 21cmFirstCLASS} the 21-cm power spectrum, given by
\begin{equation}\label{eq: 28}
\Delta_{21}^2\left(k,z\right)=\frac{k^3\bar T_{21}^2\left(z\right)P_{21}\left(k,z\right)}{2\pi^2},
\end{equation}
where $\bar T_{21}$ is the global brightness temperature and $P_{21}\left(k,z\right)$ is the angle-averaged Fourier transform of the two-point correlation function $\langle\delta_{21}\left(\mathbf x,z\right)\delta_{21}\left(\mathbf x',z\right)\rangle$, while $\delta_{21}$ is the local contrast in the brightness temperature, $\delta_{21}\left(\mathbf x,z\right)\equiv T_{21}\left(\mathbf x,z\right)/\bar T_{21}\left(z\right)-1$. We use the {\tt powerbox}\footnote{\href{https://github.com/steven-murray/powerbox}{github.com/steven-murray/powerbox}} package~\cite{2018JOSS....3..850M} to compute $\Delta_{21}^2\left(k,z\right)$ from chunks of the lightcone box of {\tt 21cmFirstCLASS}.

\begin{figure}
\includegraphics[width=0.95\columnwidth]{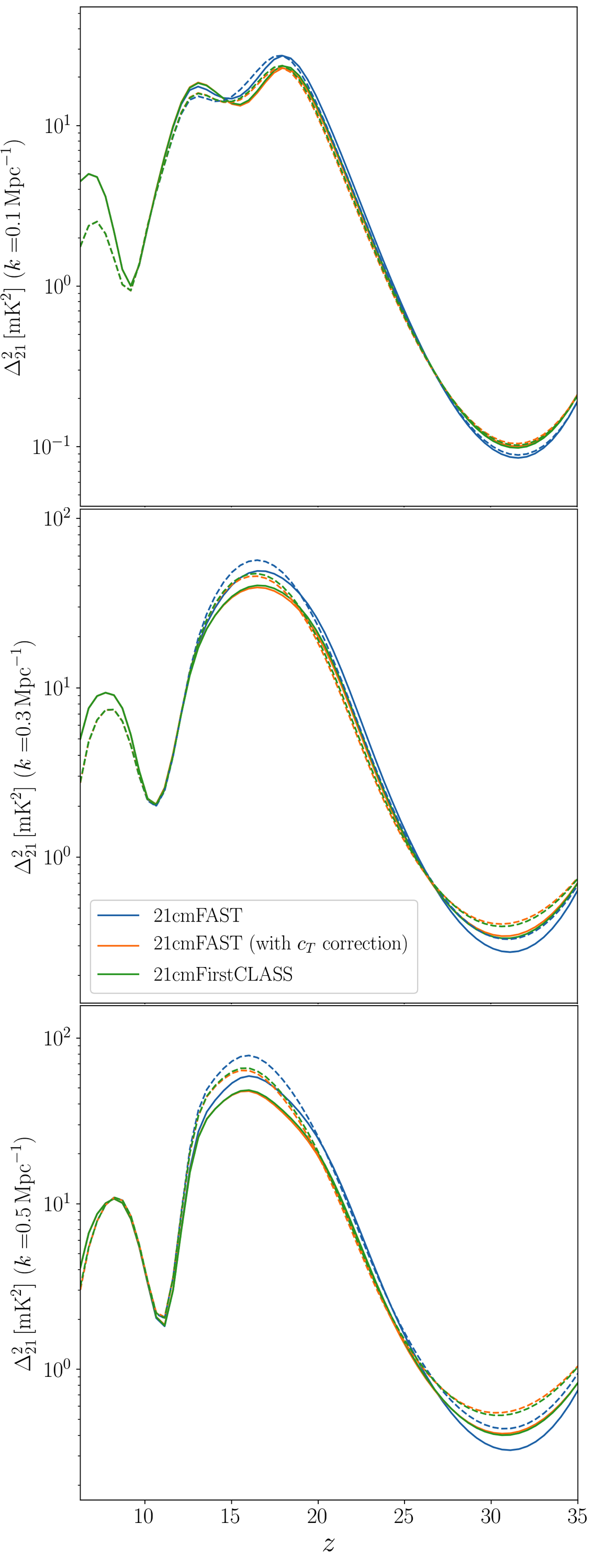}
\caption{Comparison of the 21-cm power spectrum between {\tt 21cmFirstCLASS} and {\tt 21cmFAST}, for three different wavenumbers. The blue (orange) curves show the output of {\tt 21cmFAST} without (with) the $c_T$ correction, given by the fit of JBM23 (Eq.~\ref{eq: 27}). The green curve is the output of {\tt 21cmFirstCLASS}. Solid (dashed) curves correspond to evolving the density field with 2LPT (linearly). See more details on 2LPT in Sec.~\ref{sec: non-linearities}.}
\label{Fig: figure_5}
\end{figure}

Our conclusions are shown in Fig.~\ref{Fig: figure_5}. The blue curves represent the output of the standard {\tt 21cmFAST}, where homogeneous $T_k$ and $x_e$ boxes are assumed at $z=35$. In comparison, we also plot  the 21-cm power spectrum of {\tt 21cmFirstCLASS} (which begins with homogeneous boxes at $z_\mathrm{rec}\!=\!1069$). At $25\!\lesssim\! z\!\lesssim\!35$, the early fluctuations lead to an enhancement of $\mathcal O\left(10\%\right)$. In Sec.~\ref{sec: The 21cm signal during dark ages} we will understand the origin of this enhancement. However, as we enter the non-linear regime, the opposite trend can be seen at $12\!\lesssim\! z\!\lesssim\! 25$, where early fluctuations cause a suppression of $\mathcal O\left(20\%\right)$ in the power spectrum. Afterwards, at $z\!\lesssim\!12$, the highly non-linear fluctuations of the star formation rate density (SFRD) become dominant and the initial conditions of the box at $z\!=\!35$ are forgotten.

We also show in Fig.~\ref{Fig: figure_5} the 21-cm power spectrum of the updated version of {\tt 21cmFAST}, where the fit of JBM23 (Eq.~\ref{eq: 27}) has been implemented in the code to generate an inhomogeneous $T_k$ box at $z=35$. We see that the fit of JBM23 captures very well the power spectrum of {\tt 21cmFirstCLASS}, even though this fit is slightly different than ours, Eq.~\eqref{eq: 26}. Moreover, it is interesting that the 21-cm power spectrum is insensitive to early fluctuations in $x_e$. We will understand the reason for this in Sec.~\ref{sec: The 21cm signal during dark ages}.

\subsection{Scale dependence}\label{subsec: scale dependence}

\begin{figure}
\includegraphics[width=\columnwidth]{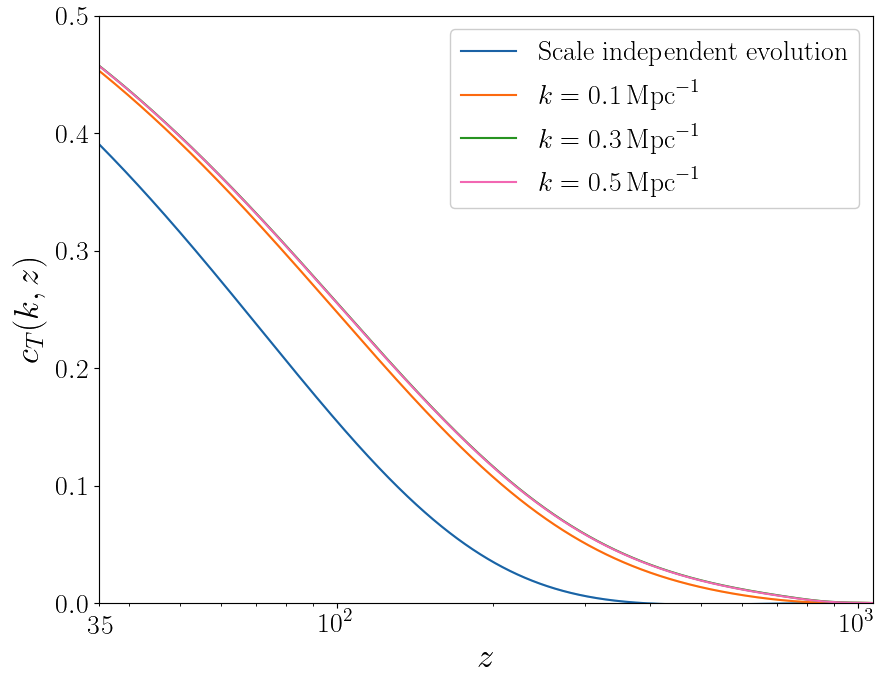}
\caption{The scale-\emph{dependent} $c_T\left(k,z\right)$. For comparison, we show the scale-\emph{indpendent} $c_T\left(z\right)$ with a blue curve (identical to the blue curve in Fig.~\ref{Fig: figure_3}). The green curve of $k=0.3\,\mathrm{Mpc}^{-1}$ completely overlaps with the pink curve of $k=0.5\,\mathrm{Mpc}^{-1}$.}
\label{Fig: figure_6}
\end{figure}

\begin{figure}
\includegraphics[width=\columnwidth]{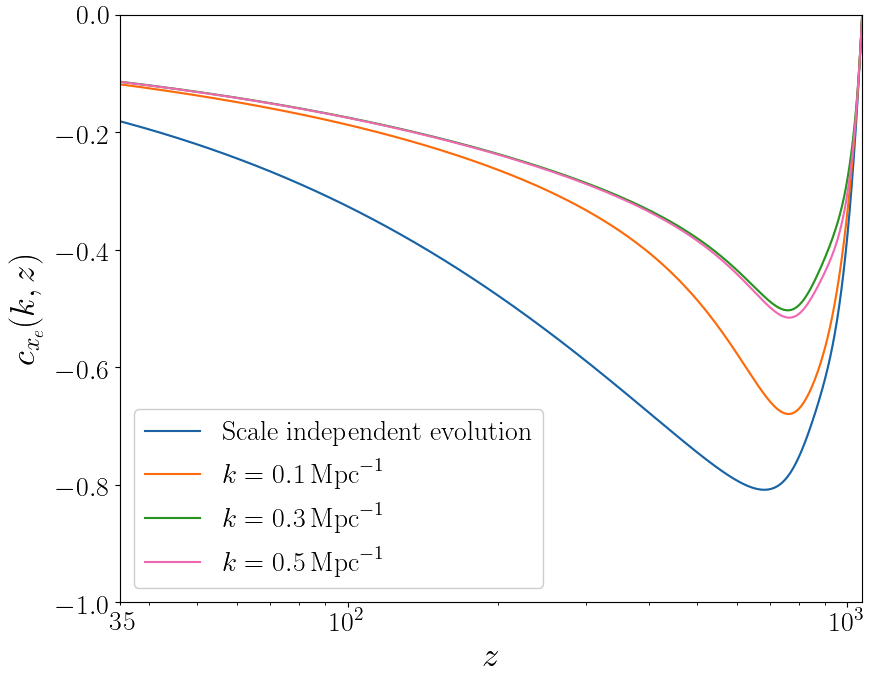}
\caption{The scale-\emph{dependent} $c_{x_e}\left(k,z\right)$. For comparison, we show the scale-\emph{indpendent} $c_{x_e}\left(z\right)$ with a blue curve (identical to the blue curve in Fig.~\ref{Fig: figure_4}).}
\label{Fig: figure_7}
\end{figure}

Although Figs.~\ref{Fig: figure_3} and \ref{Fig: figure_4} show a nice agreement between {\tt 21cmFirstCLASS} and the scale-independent theory, the assumption of $\mathcal T_{b/c}\left(k,z\right)\equiv1$ has to be relaxed if we wish to find the correct $c_T$ and $c_{x_e}$. We do that now. Using our analytical formalism above, we solve for $c_T\left(k,z\right)$ and $c_{x_e}\left(k,z\right)$, focusing on scales relevant for 21-cm interferometers. The results are shown in Figs.~\ref{Fig: figure_6} and \ref{Fig: figure_7}.

Let us analyze first how $c_{x_e}$ depends on scale. From Eq.~\eqref{eq: 21}, we can see that the source term of $\delta x_e$ is suppressed when $\mathcal D_b\left(k,z\right)$ is considered instead of the scale-independent growth factor $D\left(z\right)$. Thus, we can expect that on scales for which $\mathcal T_{b/c}\left(k,z\right)$ is smaller, the amplitude of $\delta x_e$ and $c_{x_e}$ will become smaller. This is the qualitative feature that can be seen in Figs.~\ref{Fig: figure_2} and \ref{Fig: figure_7}.

For $c_T$ on the other hand, the scale-dependence changes it in the opposite manner---scales in which $\mathcal T_{b/c}\left(k,z\right)$ is smaller lead to a higher $c_T$. This counter-intuitive behavior can be traced back to Eq.~\eqref{eq: 12}; the terms in the second and the third rows, which tend to decrease the value of $c_T$, are now smaller\footnote{Although $\mathcal D_b\left(k,z\right)$ appears both in the numerator and in the denominator of Eq.~\eqref{eq: 12}, the impact of $\mathcal T_{b/c}\left(k,z\right)$ in the numerator is stronger. This is because the numerator is inside the integral, and thus suffers more suppression from high redshifts.}. Thus, even-though the source term which drives $c_T$ to higher values is also decreased, $c_T$ is increased compared to the scale-free analysis. Specifically, for $k=0.3\,\mathrm{Mpc}^{-1}$ we find the following linear fit,
\begin{equation}\label{eq: 29}
c_T\left(k=0.3\,\mathrm{Mpc}^{-1},z\right)\approx0.46-0.0044\left(z-35\right).
\end{equation}

\section{The 21cm signal during dark ages}\label{sec: The 21cm signal during dark ages}

Because {\tt 21cmFirstCLASS} can be initialized at recombination, it can be used to calculate $\Delta_{21}^2\left(k,z\right)$ both in the non-linear cosmic dawn epoch, like {\tt 21cmFAST}, and in the dark ages. During the dark ages, the fluctuations are expected to be small, and linear perturbation theory can be applied\footnote{Though non-linear effects can contribute at the few-percent level~\cite{Lewis:2007kz, Ali-Haimoud:2013hpa}.}. This approach was taken in LC07, where the fully relativistic Boltzmann equations were solved. In order to test the output of {\tt 21cmFirstCLASS} and compare with their work, we re-derive the expressions for the linear brightness temperature fluctuations, but in a more tractable way. We do this step-by-step and identify the most dominant terms. In the derivation we present below, we make some simplifying assumptions that  help us to get a nice, short, approximated expression for the 21-cm fluctuations, Eq.~\eqref{eq: 42}. Our full derivation, where these assumptions are relaxed, can be found in Appendices \ref{sec: Brightness temperature fluctuations at the dark ages} and \ref{sec: Disentangling Ts and xCMB}. The only differences between our full derivation and the derivation of LC07 are that: (1) we ignore CMB temperature fluctuations, i.e.\ we assume $T_\gamma=\bar T_\gamma$ as fluctuations in the CMB temperature are much smaller than $\delta_b$ on subhorizon scales after decoupling; and (2) we ignore derivatives in the gravitational potential, which is justified during matter domination~\cite{Dodelson:2003ft, baumann_2022}.

First, let us assume for simplicity that there are no fluctuations in $T_k$ and $x_e$, i.e. $\delta T_k =\delta x_e =0$. This assumption is justified, at least at low redshifts, because we know from Fig.~\ref{Fig: figure_5} that the fluctuations in these fields barely change the 21-cm power spectrum at the end of the dark ages, at $z\sim35$. We will shortly understand why. Since $\bar x_\mathrm{coll}\propto\rho_b$,  under these assumptions we have $\delta x_\mathrm{coll}/\bar x_\mathrm{coll}=\delta_b$. Before cosmic dawn, the spin temperature is given simply by
\begin{equation}\label{eq: 30}
T_s=\frac{x_\mathrm{CMB}+x_\mathrm{coll}}{x_\mathrm{CMB}/T_\gamma+x_\mathrm{coll}/T_k},
\end{equation}
and thus the fractional fluctuation in the spin temperature, in linear theory, is
\begin{equation}\label{eq: 31}
\frac{\delta T_s}{\bar T_s}=-\frac{\left(1-\bar T_k/T_\gamma\right)\bar x_\mathrm{coll}\bar x_\mathrm{CMB}}{\left(\bar x_\mathrm{CMB}+\bar x_\mathrm{coll}\right)\left(\bar x_\mathrm{CMB}\bar T_k/T_\gamma +\bar x_\mathrm{coll}\right)}\delta_b.
\end{equation}
The minus sign on the RHS of Eq.~\eqref{eq: 31} makes sense, as in overdense regions collisions are more efficient and the spin temperature is driven to lower values.

In the limit $\tau_{21}\ll1$, the brightness temperature is
\begin{equation}\label{eq: 32}
T_{21}\propto\left(1-x_e\right)n_\mathrm{H}\left(1-\frac{T_\gamma}{T_s}\right)\left(1+\frac{1}{H}\frac{d\left(\mathbf{\hat n}\cdot\mathbf{v}_b\right)}{d\left(\mathbf{\hat n}\cdot\mathbf{x}\right)}\right)^{-1},
\end{equation}
where the factor of proportionality contains terms that are uniform in space. Assuming $\delta x_e=0$, we then find that the fractional fluctuation in the brightness temperature in linear theory is
\begin{equation}\label{eq: 33}
\delta_{21}\equiv\frac{\delta T_{21}}{\bar T_{21}}=\delta_b-\frac{\delta T_s}{\bar T_s}\frac{1}{1-\bar T_s/T_\gamma}-\frac{1}{H}\frac{d\left(\mathbf{\hat n}\cdot\mathbf{v}_b\right)}{d\left(\mathbf{\hat n}\cdot\mathbf{x}\right)}.
\end{equation}
Then, using Eqs.~\eqref{eq: 30}, \eqref{eq: 31} and \eqref{eq: 33}, it is straightforward to show that
\begin{equation}\label{eq: 34}
\delta_{21}\equiv c_\mathrm{21,iso}\delta_b-\frac{1}{H}\frac{d\left(\mathbf{\hat n}\cdot\mathbf{v}_b\right)}{d\left(\mathbf{\hat n}\cdot\mathbf{x}\right)},
\end{equation}
where
\begin{equation}\label{eq: 35}
c_\mathrm{21,iso}=\frac{2\bar x_\mathrm{CMB}+\bar x_\mathrm{coll}}{\bar x_\mathrm{CMB}+\bar x_\mathrm{coll}}.
\end{equation}
The physical meaning of $c_\mathrm{21,iso}$ is the ratio between $\delta_{21}$ and $\delta_b$ in the absence of peculiar velocity which introduces anisotropies in $\delta_{21}$ (see below). From the very simple expression of Eq.~\eqref{eq: 35}, we see that overdense regions lead to a stronger 21-cm signal, as expected. Moreover, the fluctuations in the signal grow in time, not only because $\delta_b$ grows, but also because $c_\mathrm{21,iso}$ grows; initially, $\bar x_\mathrm{coll}\gg \bar x_\mathrm{CMB}\approx1$ and $c_\mathrm{21,iso}\to 1$, but at late times $\bar x_\mathrm{coll}\ll \bar x_\mathrm{CMB}\approx1$ and $c_\mathrm{21,iso}\to 2$.

Eq.~\eqref{eq: 34} is a relation in real space. In Fourier space, we know from the continuity equation (assuming matter domination, when the gravitational potential was constant) that
\begin{eqnarray}\label{eq: 36}
\nonumber\mathbf{\hat n}\cdot\mathbf{v}_b&=&i\frac{\mathbf{\hat n}\cdot\mathbf{k}}{k^2}\frac{d\delta_b}{dt}=i\frac{\mathbf{\hat n}\cdot\mathbf{k}}{k^2}\frac{dz}{dt}\frac{d\delta_b}{dz}
\\&=&-iH\frac{\mathbf{\hat n}\cdot\mathbf{k}}{k^2}\frac{d\ln\mathcal T_b\left(k,z\right)}{d\ln\left(1+z\right)}\delta_b,
\end{eqnarray}
and thus in Fourier space Eq.~\eqref{eq: 34} becomes
\begin{equation}\label{eq: 37}
\delta_{21}=\left[c_\mathrm{21,iso}-\frac{\left(\mathbf{\hat n}\cdot\mathbf{k}\right)^2}{k^2}\frac{d\ln\mathcal T_b\left(k,z\right)}{d\ln\left(1+z\right)}\right]\delta_b.
\end{equation}
From here, we see that the peculiar velocity results in $c_{21}\equiv\delta_{21}/\delta_b$, which is not only scale-dependent, but also non-isotropic due to the line-of-sight effect. The 21-cm \emph{anisotropic} power spectrum can then be given in terms of the primordial curvature power spectrum $\Delta_{\mathcal R}^2\left(k\right)=A_s\left(k/k_\star\right)^{n_s-1}$,
\begin{eqnarray}\label{eq: 38}
\nonumber\Delta_{21}^2\left(k,\mu,z\right)&=&\Delta_{\mathcal R}^2\left(k\right)\bar T_{21}^2\left(z\right)c_\mathrm{21,iso}^2\left(k,z\right)\mathcal T_b^2\left(k,z\right)
\\&&\times\left[1-\mu^2c_\mathrm{21,iso}^{-1}\frac{d\ln\mathcal T_b\left(k,z\right)}{d\ln\left(1+z\right)}\right]^2,
\end{eqnarray}
where $\mu\equiv\left(\mathbf{\hat n}\cdot\mathbf{k}\right)/k$. Finally, the \emph{isotropic} (or angle-averaged) 21-cm power spectrum is
\begin{flalign}\label{eq: 39}
\nonumber&\Delta_{21}^2\left(k,z\right)=\frac{1}{4\pi}\int d\Omega\,\Delta_{21}^2\left(k,\mu,z\right)=\frac{1}{2}\int_{-1}^1 d\mu\,\Delta_{21}^2\left(k,\mu,z\right)&
\\\nonumber&=\Delta_{\mathcal R}^2\left(k\right)\bar T_{21}^2\left(z\right)c_\mathrm{21,iso}^2\left(k,z\right)\mathcal T_b^2\left(k,z\right)\times&
\\&\left[1-\frac{2}{3}c_\mathrm{21,iso}^{-1}\frac{d\ln\mathcal T_b\left(k,z\right)}{d\ln\left(1+z\right)}+\frac{1}{5}c_\mathrm{21,iso}^{-2}\left(\frac{d\ln\mathcal T_b\left(k,z\right)}{d\ln\left(1+z\right)}\right)^2\right].&
\end{flalign}
The second row in Eq.~\eqref{eq: 39} can be viewed as the 21-cm power spectrum in the absence of peculiar velocity, while the third row represents its contribution.
Let us estimate the logarithmic derivative of $\mathcal T_b\left(k,z\right)$ that appears in the third row,
\begin{eqnarray}\label{eq: 40}
\nonumber\frac{d\ln\mathcal T_b\left(k,z\right)}{d\ln\left(1+z\right)}&=&\frac{d\ln\mathcal T_c\left(k,z\right)}{d\ln\left(1+z\right)}+\frac{d\ln\mathcal T_{b/c}\left(k,z\right)}{d\ln\left(1+z\right)}
\\&=&-1+\frac{d\ln\mathcal T_{b/c}\left(k,z\right)}{d\ln\left(1+z\right)}<0,
\end{eqnarray}
where the second equality follows $\mathcal T_c\left(k,z\right)\propto D\left(z\right)\propto\left(1+z\right)^{-1}$ and the inequality is due to the fact that the slope of $\mathcal T_{b/c}\left(k,z\right)$ is non-positive; from Fig.~\ref{Fig: figure_2} we see that this derivative is almost zero at low redshifts, while at high redshifts it can be $\mathcal O\left(-10\right)$. Thus, we see from Eqs.~\eqref{eq: 39}-\eqref{eq: 40} that the peculiar velocity tends to add power to the 21-cm signal. If we ignore the logarithmic derivative of $\mathcal T_{b/c}\left(k,z\right)$ and adopt the expression for $c_\mathrm{21,iso}$ from Eq.~\eqref{eq: 35}, then we see from Eq.~\eqref{eq: 39} that at high redshifts the peculiar velocity enhances $\Delta_{21}^2\left(k,z\right)$ by a factor of $1+\frac{2}{3}+\frac{1}{5}=\frac{28}{15}=1.87$~\cite{Lewis:2007kz, Jensen:2013fha, Mao:2011xp}, while at low redshifts it enhances $\Delta_{21}^2\left(k,z\right)$ by a factor of $1+\frac{1}{3}+\frac{1}{20}=\frac{83}{60}=1.38$.

\begin{figure}
\includegraphics[width=\columnwidth]{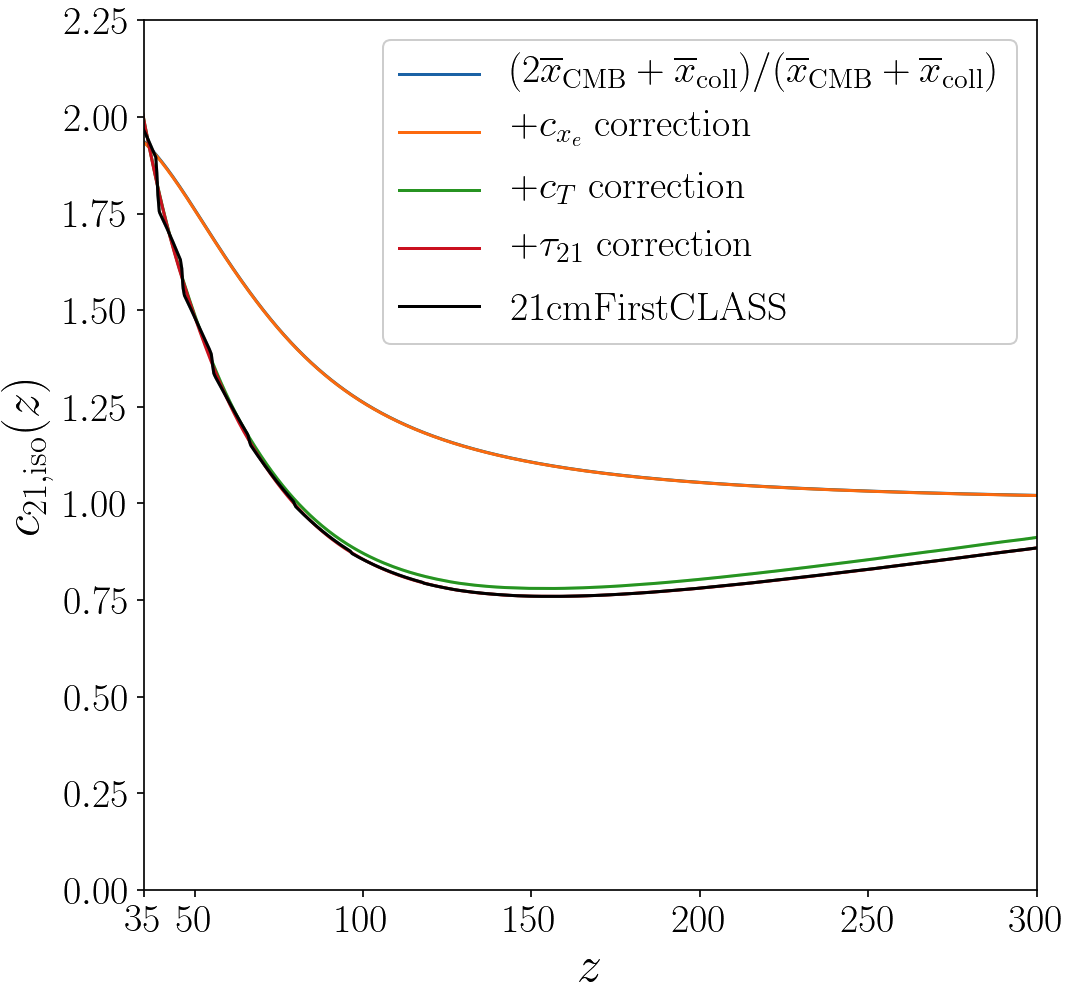}
\vspace{-0.25in}
\caption{The scale-\emph{independent} $c_\mathrm{21,iso}\left(z\right)$. The blue curve shows our crude approximation, Eq.~\eqref{eq: 35}. The orange curve shows the contribution from early ionization fluctuations, the first and the second rows in Eq.~\eqref{eq: 41}, and it completely overlaps with the blue curve. The green curve shows the contribution from early temperature fluctuations, which is achieved by considering all the rows in Eq.~\eqref{eq: 41}, or alternatively via Eq.~\eqref{eq: 42} with excellent precision. The red curve shows the contribution from $\tau_{21}$ fluctuations, see details in Appendices~\ref{sec: Brightness temperature fluctuations at the dark ages} and \ref{sec: Disentangling Ts and xCMB}. The black curve shows the redshift-dependent median of the $c_{21}\left(\mathbf x,z\right)$ box in {\tt 21cmFirstCLASS} (when the peculiar velocity and {\tt HyRec} are turned off) and it completely overlaps with the red curve at high redshifts. At low redshifts, the jagged artifact is the result of the linear interpolation used in {\tt 21cmFAST}, see text for more details.}
\label{Fig: figure_8}
\end{figure}

Although Eq.~\eqref{eq: 39} is theoretically correct\footnote{Neglecting $\mathcal O\left(\tau_{21}\right)$ corrections in the third row, see Appendices~\ref{sec: Brightness temperature fluctuations at the dark ages} and \ref{sec: Disentangling Ts and xCMB}.}, the expression we derived for $c_\mathrm{21,iso}$ in Eq.~\eqref{eq: 35} is too simplistic and lacks the contribution of fluctuations in $T_k$ and $x_e$. Once these fluctuations are taken into account, the expression for $c_\mathrm{21,iso}$ becomes (still assuming $\tau_{21}\ll1$, see details in Appendix~\ref{sec: Brightness temperature fluctuations at the dark ages})
\begin{widetext}
\begin{eqnarray}\label{eq: 41}
\nonumber c_\mathrm{21,iso}\left(k,z\right)&=&\frac{2\bar x_\mathrm{CMB}+\bar x_\mathrm{coll}}{\bar x_\mathrm{CMB}+\bar x_\mathrm{coll}}
\\\nonumber&&+\left[\frac{\bar x_\mathrm{CMB}\bar C_{10}^{-1}\bar n_\mathrm{H}\bar x_e}{\bar x_\mathrm{CMB}+\bar x_\mathrm{coll}}\left(f_\mathrm{H}^{-1}\bar \kappa^\mathrm{eH}_{1-0}+\bar \kappa^\mathrm{pH}_{1-0}-\bar \kappa^\mathrm{HH}_{1-0}\right)+\frac{\bar x_e}{1-\bar x_e}\right]c_{x_e}\left(k,z\right)
\\&&+\left[\frac{\bar x_\mathrm{CMB}\bar C_{10}^{-1}\bar n_\mathrm{H}\bar T_k}{\bar x_\mathrm{CMB}+\bar x_\mathrm{coll}}\left(\left(1-\bar x_e\right)\frac{\partial\bar\kappa^\mathrm{HH}_{1-0}}{\partial\bar T_k}+f_\mathrm{H}^{-1}\bar x_e\frac{\partial\bar\kappa^\mathrm{eH}_{1-0}}{\partial\bar T_k}+\bar x_e\frac{\partial\bar\kappa^\mathrm{pH}_{1-0}}{\partial\bar T_k}\right)-\frac{1}{1-\bar T_k/T_\gamma}\right]c_T\left(k,z\right).
\end{eqnarray}
\end{widetext}
Note that since $c_T$ and $c_{x_e}$ are scale-dependent quantities, so too is $c_\mathrm{21,iso}$. In addition, note that the contribution of $c_{x_e}$ to $c_\mathrm{21,iso}$ is proportional to $\bar x_e$, which is very small between recombination and reionization, and we can therefore take the approximation $\bar x_e\approx0$ in Eq.~\eqref{eq: 41},
\begin{flalign}\label{eq: 42}
\nonumber& c_\mathrm{21,iso}\left(k,z\right)\approx\frac{2\bar x_\mathrm{CMB}+\bar x_\mathrm{coll}}{\bar x_\mathrm{CMB}+\bar x_\mathrm{coll}}&
\\&\hspace{10mm}+\left[\frac{\bar x_\mathrm{CMB}}{\bar x_\mathrm{CMB}+\bar x_\mathrm{coll}}\frac{\partial\ln\bar\kappa^\mathrm{HH}_{1-0}}{\partial\ln\bar T_k}-\frac{1}{1-\bar T_k/T_\gamma}\right]c_T\left(k,z\right).&
\end{flalign}
Although this expression may still appear rather simplistic, we stress that it results in only a sub-percent error below $z\lesssim80$ (above this redshift, corrections from the optical depth become more important, but are still not very significant, see Fig.~\ref{Fig: figure_8}). Together with Eqs.~\eqref{eq: 38}-\eqref{eq: 40}, it can therefore  be used to gain insight as to how the 21-cm power spectrum behaves during the dark ages.

Notice that if the expression in the square brackets of Eq.~\eqref{eq: 42} vanishes, then $c_\mathrm{21,iso}$ does not depend on early temperature fluctuations, and it becomes completely  scale-independent. We find that the special redshift where that happens is $z\approx37$. This explains why the power spectrum in {\tt 21cmFAST} is not very sensitive to early temperature fluctuations at the vicinity of $z=35$, as was demonstrated in Fig.~\ref{Fig: figure_5}. At lower redshifts, the weight of $c_T$ becomes positive and larger, and since its evolution depends on its past values, the initial conditions at $z=35$ carry more importance, until the non-linearities of the SFRD dominate the power spectrum.

In Fig.~\ref{Fig: figure_8} we show how each of the lines in Eq.~\eqref{eq: 41} contributes to $c_\mathrm{21,iso}\left(z\right)$, including the $\mathcal O\left(\tau_{21}\right)$ term that we have neglected above, while taking the scale-\emph{independent} quantities of $c_T\left(z\right)$ and $c_{x_e}\left(z\right)$. As expected, the crude approximation of Eq.~\eqref{eq: 35} works well at low redshifts, at $z\sim35$. The $c_{x_e}$ correction barely modifies $c_\mathrm{21,iso}$, while the $c_T$ correction modifies $c_\mathrm{21,iso}$ by $\sim30\%$ at $z\sim100$ and its impact becomes smaller towards lower redshifts. The $\tau_{21}$ correction is mainly important at high redshifts, $z\gtrsim100$ ($\bar \tau_{21}$ monotonically decreases from $\sim0.08$ at $z\sim700$ to $\sim0.02$ at $z\sim35$). We note that we cut Fig.~\ref{Fig: figure_8} at $z=300$ because of a theoretical uncertainty arising from the last term in Eq.~\eqref{eq: 42}; while we know that $c_T$ becomes very small at high redshifts, this is the regime where $\bar T_k\to T_\gamma$ and it is not clear which one approaches  zero more quickly, the numerator or the denominator.

As a sanity check, we also plot in Fig.~\ref{Fig: figure_8} the median of the $c_{21}\left(\mathbf x,z\right)$ box in {\tt 21cmFirstCLASS}. For this test we have turned off the peculiar velocity in {\tt 21cmFAST} such that $c_{21}\equiv c_\mathrm{iso,21}$. We see again that the agreement between {\tt 21cmFirstCLASS} and the linear scale-independent theory is excellent at high redshifts. Below $z\sim80$ the black curve becomes jagged. This feature is of course unphysical and it is the result of the evaluation of $\kappa^\mathrm{HH}_{1-0}\left(T_k\right)$ in {\tt 21cmFAST}. This quantity (as well as $\kappa^\mathrm{pH}_{1-0}$ and $\kappa^\mathrm{eH}_{1-0}$) is \emph{linearly} interpolated from an external table. According to Eq.~\eqref{eq: 42}, the derivative of $\kappa^\mathrm{HH}_{1-0}$ with respect to $T_k$ has to be considered, and thus the jagged artifact\footnote{In Fig.~\ref{Fig: figure_8} we have evaluated all the colorful curves by using {\tt Scipy}'s intepolation function with a cubic interpolation scheme~\cite{2020SciPy-NMeth}. We have replicated the jagged artifact shown in the black curve by using instead a linear interpolation scheme.} is the consequence of {\tt 21cmFirstCLASS} evaluating $\delta\kappa^\mathrm{HH}_{1-0}/\delta T_k$ from a piecewise linear function. Because this artifact has no effect on the results shown in this paper, we defer its fix for future work. 

\section{Non-linearities}\label{sec: non-linearities}

Unlike the linear Boltzmann solver {\tt CAMB}, {\tt 21cmFirstCLASS} can take into account non-linear fluctuations that arise from non-linear growth of the density field. Below we outline the Lagrangian perturbation theory (LPT)~\cite{1978ApJ...221..937F, Buchert:1989xx, Moutarde:1991evx, 1993A&A...267L..51B, Hivon:1994qb, Carlson:2009it, Tatekawa:2004mq} scheme used in {\tt 21cmFAST} to evolve the density field non-linearly, but we generalize our formulae to account for scale-dependent growth.

The basic idea is to relate the initial (comoving) Lagrangian coordinates $\mathbf x$ of the density field, evaluated linearly at some high redshift $z_0$ with the Eulerian coordinates $\mathbf x_E$ of the density field at a lower redshift $z$. These two coordinates are related by the displacement field vector $\boldsymbol\psi\left(\mathbf x,z\right)$,
\begin{equation}\label{eq: 43}
\mathbf x_E\left(\mathbf x,z\right)=\mathbf x\left(z_0\right)+\boldsymbol\psi\left(\mathbf x,z\right).
\end{equation}
The information encoded in $\boldsymbol\psi\left(\mathbf x,z\right)$ maps the spatial displacement between redshift $z_0$ and redshift $z$ of a mass located initially at coordinate $\mathbf x$. Knowing the value of $\boldsymbol\psi\left(x,z\right)$ then allows to coherently add masses together at $\mathbf x_E$, resulting in non-linear fluctuations at sufficiently low redshifts. By construction, the time derivative of $\boldsymbol\psi\left(\mathbf x,z\right)$ is the velocity field, $\dot{\boldsymbol\psi}\left(\mathbf x,z\right)\!=\!\dot{\mathbf x}_E\left(\mathbf x,z\right)\!=\!\mathbf v_b\left(\mathbf x,z\right)$, where we use the subscript $b$ to emphasize that we are focusing on the baryon velocity. In the Zel'dovich approximation~\cite{1970A&A.....5...84Z, White:2014gfa, Hidding:2013kka}, the velocity of the baryons can be expressed with the baryons density contrast $\delta_b$ via the continuity equation, which in Fourier space reads~\cite{1985ApJS...57..241E, Sirko:2005uz,Mesinger:2007pd}
\begin{equation}\label{eq: 44}
\mathbf v_b\left(\mathbf k,z\right)=\frac{i\mathbf k}{k^2}\dot\delta_b\left(\mathbf k,z\right)=\frac{i\mathbf k}{k^2}\dot{\mathcal D}_b\left(k,z\right)\delta_0\left(\mathbf k\right).
\end{equation}
Thus, in the Zel'dovich approximation, the displacement field vector in Fourier space is given by
\begin{equation}\label{eq: 45}
\boldsymbol\psi\left(\mathbf k,z\right)=\frac{i\mathbf k}{k^2}\left[\mathcal D_b\left(k,z\right)-\mathcal D_b\left(k,z_0\right)\right]\delta_0\left(\mathbf k\right).
\end{equation}
Note that unlike the scale-independent scenario, here $\boldsymbol\psi\left(\mathbf k,z\right)$ cannot be separated into its spatial and temporal components.

Despite the fact that the Zel'dovich approximation correctly reproduces the linear growth of density and velocity perturbations, it fails to conserve momentum in the non-linear regime. Therefore, second order Lagrangian perturbation theory (2LPT) is implemented in {\tt 21cmFAST}. The basic equations of 2LPT can be found in Appendix D of Ref.~\cite{Scoccimarro:1997gr} and the references therein. These equations are expressed in terms of the scale-independent growth factor $D\left(z\right)$, and the scale-dependence can be accounted by replacing $D\left(z\right)$ with $\mathcal D_b\left(k,z\right)$, as we did above for the Zel'dovich approximation.

For the time being, as we have yet to incorporate the scale-dependent growth factor $\mathcal D_b\left(k,z\right)$ in {\tt 21cmFirstCLASS}, we can currently study how the non-linear evolution scheme of 2LPT modifies the analytical results we previously obtained for the scale-independent scenario. In that context, we have already seen in Fig.~\ref{Fig: figure_5} the effect of non-linearities at low redshifts on the 21-cm power spectrum. On large scales ($k=0.1\,\mathrm{Mpc}^{-1}$) non-linearities enhance the power spectrum by $\mathcal O\left(100\%\right)$ at $z\sim7$, while on small scales ($k=0.5\,\mathrm{Mpc}^{-1}$) they suppress the power spectrum by $\mathcal O\left(25\%\right)$. This will be revisited in follow-up work.

\section{Comparison with {\tt CAMB}}\label{sec: Comparison with CAMB}

So far, we have compared our scale-independent linear theory with {\tt 21cmFirstCLASS} and we have generalized the equations to account for scale-dependent growth by simply replacing $D\left(z\right)$ with $\mathcal D_b\left(k,z\right)$. The key question that emerges now is the following: once scale-dependent growth is implemented in {\tt 21cmFirstCLASS}, will it be able to predict the 21-cm anisotropies during the dark ages?

Currently, the code that computes the linear 21-cm anisotropies during the dark ages most comprehensively is {\tt CAMB}. Similarly to {\tt CLASS}, this code solves the fully-relativistic coupled Boltzmann and Einstein equations for the linear scale-dependent perturbations in the density, velocity and temperature fields. Once {\tt CAMB} has evaluated these perturbations, it can compute the fluctuations in the spin temperature and the brightness temperature with very similar equations to those derived in Sec.~\ref{sec: The 21cm signal during dark ages} (except that CAMB also accounts for fluctuations in the CMB temperature and does not ignore the small impact of gravitational fields, see more details in LC07).

\begin{figure}
\includegraphics[width=\columnwidth]{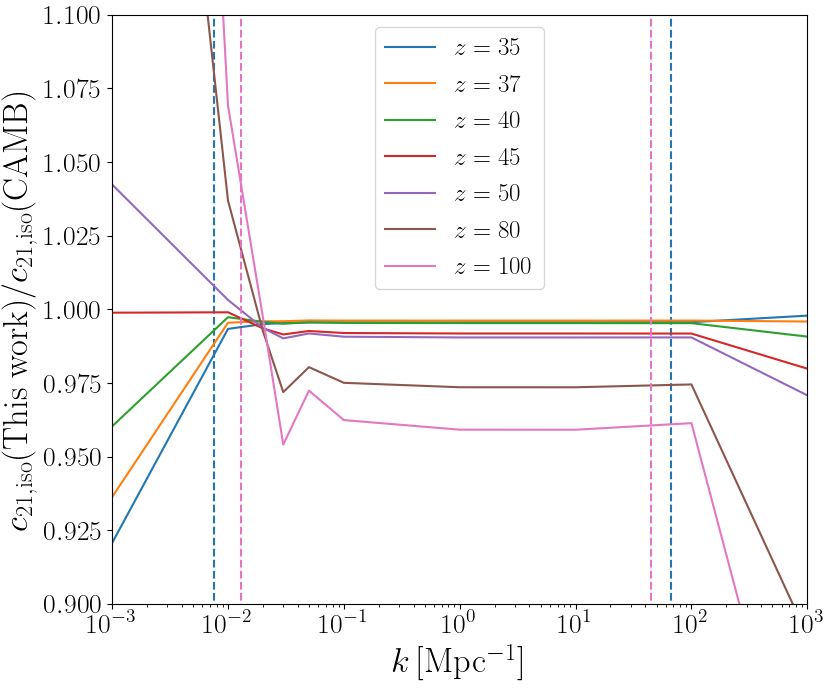}
\caption{Comparison of the scale-\emph{dependent} $c_\mathrm{21,iso}\left(k,z\right)$ between the formalism developed in this work and {\tt CAMB}, as a function of wavenumber, for various redshifts. The vertical dashed curves represent the wavenumbers associated with the horizon scale (times 10) and the Jeans scale (divided by 10).}
\label{Fig: figure_10}
\end{figure}

Because the scale-dependent formalism we developed in this work is less rigorous than what {\tt CAMB} does, it is not a priori clear that our formalism gives the correct results. We therefore present a comparison with {\tt CAMB} in Fig.~\ref{Fig: figure_10} for a wide range of scales between $35\leq z\leq100$. The vertical dashed curves represent the wavenumbers associated with the horizon scale (times 10) and the Jeans scale (divided by 10). In the intermediate scales that span between these lines the comparison is very successful; for $z\leq40\,,50,\,100$ the relative error is less than $0.5,1,5\%$, respectively. As expected, our formalism fails to predict the 21-cm anisotropies on superhorizon scales, since we have neglected in our equations the fluctuations in the CMB temperature. On scales smaller than the Jeans scale, the discrepancy is less severe, but still exists because at these small scales the evolution equations of $\delta_b$ and $\delta T_k$ are coupled~\cite{Naoz:2005pd, Tseliakhovich:2010yw}, and we cannot simply take $\mathcal D_b\left(k,z\right)$ from {\tt CLASS} to emulate the evolution of the baryon density field. Plus, Jeans suppression causes $\delta_b$ to approach $\delta_\gamma$, the density contrast of the CMB photons which we have neglected. 

A very nice feature that can be seen in Fig.~\ref{Fig: figure_10} is that on subhorizon scales, the relative error is constant at $z=37$, confirming our discussion below Eq.~\eqref{eq: 42} that $c_\mathrm{21,iso}$ becomes scale-independent at this special redshift. Another feature that can be noticed is that the relative error increases with redshift. We suspect that the origin of this feature is that {\tt CAMB} does not only evaluate the fluctuations differently than us, it also computes the background quantities differently. For example, throughout this paper we have assumed that the spin temperature is given by its expression in equilibrium, Eq.~\eqref{eq: 3}. Yet, {\tt CAMB} solves the Boltzmann equation for the background spin temperature, Eq.~(28) in LC07, thereby accounting for out-of-equilibrium effects in the spin temperature. 

All this notwithstanding, it is important to realize that once the scale-dependence is implemented in {\tt 21cmFirstCLASS}, we still should not expect it to match {\tt CAMB} with less than a few percent difference. As was discussed at the end of Sec.~\ref{subsec: Scale independence}, {\tt 21cmFirstCLASS} works with the more precise recombination model of {\tt HyRec}, while {\tt CAMB} works with {\tt RECFAST}. As we demonstrated in Fig.~\ref{Fig: figure_3}, the different recombination models lead to a relative difference of $\sim4\%$ in $c_T$ at $z=100$, and according to Fig.~\ref{Fig: figure_8}, the weight of $c_T$ in determining $c_\mathrm{21,iso}$ becomes important at that redshift.

Moreover, {\tt 21cmFirstCLASS} includes two effects\footnote{Another effect, which is absent in both {\tt 21cmFirstCLASS} and {\tt CAMB}, is the inclusion of super-sonic relative velocity between baryons and CDM~\cite{Tseliakhovich:2010bj, Tseliakhovich:2010yw, Fialkov:2011iw}. This is a non-linear effect arising from very small scales ($k\gtrsim100\,\mathrm{Mpc}^{-1}$) and can enhance the 21cm power spectrum at the relevant scales ($0.005\lesssim k\lesssim1\,\mathrm{Mpc}^{-1}$) by a few percent~\cite{Ali-Haimoud:2013hpa}.} that {\tt CAMB} does not in the evaluation of $\Delta_{21}^2\left(k,z\right)$, both of them were first included in {\tt 21cmFAST} in Ref.~\cite{Greig:2018hja}. The first one is redshift space distortions (RSD). The peculiar velocity impacts the brightness temperature in two ways. Firstly, local peculiar velocity gradients modify the amplitude of the brightness temperature (this effect is already included in {\tt CAMB} and is captured by our Eqs.~\ref{eq: 1} and \ref{eq: 2}). And secondly, because observations are performed in redshift space rather than real space, the signal shifts from real space coordinates $\mathbf x$ to redshift space coordinates $\mathbf s$ via $\mathbf s=\mathbf x+\mathbf{\hat n}\left(\mathbf{\hat n}\cdot\mathbf v_b\right)/H$~\cite{Ali-Haimoud:2013hpa, Jensen:2013fha, Mao:2011xp}. In order to simulate this effect, cells in the box of {\tt 21cmFAST} are shifted accordingly. The second effect that is already accounted in {\tt 21cmFirstCLASS} is the lightcone effect~\cite{McQuinn:2005hk, Barkana:2005jr, Zawada:2014jea, Datta:2011hv, Datta:2014fna, LaPlante:2013jhi, Ghara:2015yfa, Mondal:2017qsq}---measurements of modes parallel to the line-of-sight probe the brightness temperature at different redshifts. The lightcone effect is addressed in {\tt 21cmFAST} by building the lightcone box from interpolating adjacent coeval boxes (the latter are boxes for which all their cells correspond to a single redshift) along the line-of-sight. The RSD is a non-linear effect and is beyond the scope of this work (but see analytical treatment in Refs.~\cite{Ali-Haimoud:2013hpa, Mao:2011xp}). The lightcone effect is even more challenging to analytically model as it mixes between time and space coordinates (but see an analytical estimate of the effect at the beginning of reionization in Ref.~\cite{McQuinn:2005hk}). We thus leave a more thorough comparison with {\tt CAMB} for future work, once the scale-dependence is implemented in {\tt 21cmFirstCLASS}.

\section{Conclusions}\label{sec: Conclusions}

The 21-cm signal is expected to provide a new window to the thermal and ionization states of the high-redshift Universe. The rich information in that signal, especially in its anisotropies pattern, cannot be utilized without having robust theoretical models. There are two important epochs in the history of the signal, each of which is modelled nowadays by a different theoretical framework. During the dark ages, above $z\sim35$, the fluctuations in the signal are expected to be linear, hence they can be evolved via linear perturbation theory. This is what the {\tt CAMB} code does rigorously, by solving the full Einstein-Boltzmann equations for all the linear fluctuations. At low redshifts, below $z\sim35$, non-linearities due to structure formation become important and semi-numerical codes, like {\tt 21cmFAST}, are required for studying the vast cosmological and astrophysical parameter space.

In the companion Paper I, we introduced our new code, {\tt 21cmFirstCLASS}. As an extension to {\tt 21cmFAST}, it is automatically able to calculate the 21-cm power spectrum during the non-linear cosmic dawn epoch. To test the output of our code, we have developed here a new analytical framework to predict the linear contribution to the brightness temperature anisotropies. We first showed that initializing the evolution at recombination can lead to differences at the $\mathcal{O}(20\%)$ level in the cosmic dawn signal compared to the standard {\tt 21cmFAST} result. 

In addition, the equations we derived indicate that {\tt 21cmFAST} over-predicts the brightness temperature fluctuations at $z\gtrsim20$ by as much as $\mathcal O\left(10\%\right)$, which leads to a $\mathcal O\left(20\%\right)$ error in the 21-cm power spectrum at these redshifts. This arises because of a basic underlying assumption in {\tt 21cmFAST}, that the densities of baryons and cold dark matter are the same, i.e.\ $\delta_b=\delta_c$.

We then proposed an elegant solution to this problem: simply replace the scale-independent growth factor $D\left(z\right)$ with the scale-dependent growth factor $\mathcal D_b\left(k,z\right)$. Then, instead of solving dozens of coupled differential equations like {\tt CAMB}, only two are required, one for $\delta T_k$ and one for $\delta x_e$. These two equations are given in Eqs.~\eqref{eq: 15}-\eqref{eq: 21} (along with Eqs.~\eqref{eq: A7}-\eqref{eq: A9} to account for Peebles fluctuations), and their analytical solution is given by Eq.~\eqref{eq: 22}. For the brightness temperature fluctuations, only the early fluctuations in $T_k$ matter, and we have managed to derive a simple analytical formula for $c_\mathrm{21,iso}\equiv\delta_{21}/\delta_b$ (defined without the contribution of the peculiar velocity), Eq.~\eqref{eq: 42}, that works particularly well below $z\lesssim80$. This equation can then be used alongside Eqs.~\eqref{eq: 38}-\eqref{eq: 39} in the evaluation of the 21-cm dark ages power spectrum.

As a byproduct of this work, we have found that at $z\approx37$ the 21-cm power spectrum becomes scale-independent and completely independent from the history of temperature fluctuations. Below that redshift, temperature fluctuations can matter in the evolution of the brightness temperature fluctuations, as mentioned above. Their weight at redshifts $25\lesssim z\lesssim35$ is still small, however, explaining why {\tt 21cmFirstCLASS}, as well as the fit of JBM23, yields only a $\sim10\%$ difference there.

In order to verify that the formalism we have developed can be incorporated in the future in {\tt 21cmFirstCLASS} for predicting $\Delta_{21}^2\left(k,z\right)$ at the dark ages with a good precision level, our strategy was the following (see Fig.~\ref{Fig: figure_1}): First, we compared between the scale-independent theory and {\tt 21cmFirstCLASS}. The results of this comparison, shown in Fig.~\ref{Fig: figure_8}, are excellent, {\tt 21cmFirstCLASS} does exactly what it was designed to do (except for the jagged artifact at low redshifts which can be easily fixed in future versions of the code). And secondly, we compared between the scale-dependent theory and {\tt CAMB}. The results of this comparison are presented in Fig.~\ref{Fig: figure_10}. At the scales of interest for 21-cm interferometers, between the horizon and the Jeans scale, our formalism appears to be robust, and there is only a sub-percent discrepancy with {\tt CAMB} at $z\leq50$ over a wide dynamic range of scales. As was discussed above, in the end, a few percent differences between {\tt 21cmFirstCLASS} and {\tt CAMB} are unavoidable because of the different recombination models used by the two codes, as well as other effects that are accounted for in {\tt 21cmFirstCLASS} but not in {\tt CAMB} (non-linearities in the density field, RSD and the lightcone effect). 

We are therefore  optimistic that our formalism can be incorporated in {\tt 21cmFirstCLASS}, making its upcoming public release a desirable code for studying the 21-cm anisotropies not only at cosmic dawn, but also during the dark ages. Its greatest strength, though, will be in the study of non-linear models, as we demonstrate in Paper I. {\tt 21cmFirstCLASS} is thus poised to become a widely used tool for cosmological simulations of $\Lambda$CDM and beyond.

\begin{acknowledgments}
We thank Julian B. Mu\~noz for useful discussions. We also acknowledge the efforts of the  {\tt 21cmFAST} and {\tt CLASS} and {\tt CAMB} authors to produce state-of-the-art public 21-cm and CMB codes. JF is supported by the Zin fellowship awarded by the BGU Kreitmann School. EDK acknowledges support from an Azrieli faculty fellowship. EDK also acknowledges joint support from the U.S.-Israel Bi-national Science Foundation (BSF, grant No. 2022743) and the U.S. National Science Foundation (NSF, grant No. 2307354), and support from the ISF-NSFC joint research program (grant No. 3156/23).

\end{acknowledgments}

\appendix
\section{Peebles fluctuations}\label{sec: Peebles fluctuations}
The Peebles coefficient in Eq.~\eqref{eq: 6} represents the probability that a hydrogen atom in the first excited state reaches the ground state before being ionized. It is given by~\cite{1968ApJ...153....1P,Dodelson:2003ft,baumann_2022}
\begin{equation}\label{eq: A1}
\mathcal C=\frac{\Lambda_\alpha+\Lambda_{2\gamma}}{\Lambda_\alpha+\Lambda_{2\gamma}+\beta_\mathrm{ion}^{(2)}},
\end{equation}
where $\Lambda_{2\gamma}=8.227\,\mathrm{sec}^{-1}$ is the two-photon decay rate from the $2s$ level to the $1s$ level, $\Lambda_\alpha$ is the resonance escape of Ly$\alpha$ photons,
\begin{equation}\label{eq: A2}
\Lambda_\alpha=\frac{8\pi H}{\lambda_\alpha^3n_\mathrm{H}\left(1-x_e\right)},
\end{equation}
where $\lambda_\alpha=121.6\,\mathrm{nm}$ is the Ly$\alpha$ wavelength, and the photoionization rate from the first excited state is given by
\begin{equation}\label{eq: A3}
\beta_\mathrm{ion}^{(2)}=\beta_\mathrm{ion}\mathrm{e}^{3\epsilon_0/\left(4k_BT_\gamma\right)}=\alpha_\mathrm{rec}\left(\frac{m_ek_BT_\gamma}{2\pi\hbar^2}\right)^{3/2}\mathrm{e}^{-\epsilon_0/\left(4k_BT_\gamma\right)}.
\end{equation}

Applying linear perturbation theory on Eq.~\eqref{eq: A1}, one finds that
\begin{equation}\label{eq: A4}
\frac{\delta\mathcal C}{\overline{\mathcal C}}=\overline{\mathcal C}\frac{\bar \beta_\mathrm{ion}^{(2)}}{\bar\Lambda_\alpha+\Lambda_{2\gamma}}\left(-\frac{\delta\beta_\mathrm{ion}^{(2)}}{\bar \beta_\mathrm{ion}^{(2)}}+\frac{\bar\Lambda_\alpha}{\bar\Lambda_\alpha+\Lambda_{2\gamma}}\frac{\delta\Lambda_\alpha}{\bar\Lambda_\alpha}\right).
\end{equation}
From Eqs.~\eqref{eq: A2}-\eqref{eq: A3} we see that
\begin{equation}\label{eq: A5}
\frac{\delta\Lambda_\alpha}{\bar\Lambda_\alpha}=-\delta_b+\frac{\delta x_e}{1-\bar x_e}
\end{equation}
\begin{equation}\label{eq: A6}
\frac{\delta\beta_\mathrm{ion}^{(2)}}{\bar \beta_\mathrm{ion}^{(2)}}=\frac{\delta\alpha_\mathrm{rec}}{\bar\alpha_\mathrm{rec}}=\frac{\partial\ln\alpha_B}{\partial \bar T_k}\delta T_k
\end{equation}
where we assumed $\alpha_\mathrm{rec}\approx\alpha_B\left(T_k\right)$, just as in Sec.~\ref{sec: Early temperature and ionization fluctuations}. From Eqs.~\eqref{eq: 14}-\eqref{eq: 15} and Eqs.~\eqref{eq: A4}-\eqref{eq: A6} we see that the Peebles fluctuations modify the elements of the $A$ and $B$ matrices according to
\begin{eqnarray}
A_{21}&\to&A_{21}-\overline{\mathcal C}\frac{\bar \beta_\mathrm{ion}^{(2)}}{\bar\Lambda_\alpha+\Lambda_{2\gamma}}\frac{\partial\ln\alpha_B}{\partial \bar T_k}\frac{d\bar x_e}{dz}\label{eq: A7} \\
A_{22}&\to&A_{22}+\overline{\mathcal C}\frac{\bar \beta_\mathrm{ion}^{(2)}}{\left(\bar\Lambda_\alpha+\Lambda_{2\gamma}\right)^2}\frac{1}{1-\bar x_e}\frac{d\bar x_e}{dz}\label{eq: A8} \\
B_2&\to&B_2-\overline{\mathcal C}\frac{\bar \beta_\mathrm{ion}^{(2)}}{\left(\bar\Lambda_\alpha+\Lambda_{2\gamma}\right)^2}\frac{d\bar x_e}{dz}\mathcal D_b\left(k,z\right).\label{eq: A9}
\end{eqnarray}

It should be noted that in principle Eq.~\eqref{eq: A5} is missing a term proportional to $\mathbf k\cdot\mathbf v_b$~\cite{Lewis:2007kz, Ali-Haimoud:2013hpa, Senatore:2008vi}, whose origin comes from modifications in $\Lambda_\alpha$ due to local peculiar velocities. It is justified however to discard this term in our analysis because (1) Peebles fluctuations only affect $c_{x_e}$ at high redshift, leaving $c_T$ unchanged, and (2) fluctuations in the brightness temperature are mostly determined by $c_T$, not $c_{x_e}$ (see Eq.~\eqref{eq: 42}).

\section{Brightness temperature fluctuations at the dark ages}\label{sec: Brightness temperature fluctuations at the dark ages}
In this appendix we derive the linear brightness temperature fluctuations during the dark ages. This is similar to the calculation presented at Sec.~\ref{sec: The 21cm signal during dark ages} but here we relax the assumptions that were made in the main text. To make the calculation as tractable as possible, we begin by assuming $\tau_{21}\ll1$, but let the temperature and ionization fraction fluctuate.

The collision coupling is given by $x_\mathrm{coll}=T_\star C_{10}/\left(A_{10}T_\gamma\right)$, where $T_\star=68.2\,\mathrm{mK}$ is the hydrogen hyperfine energy gap (in units of mK), and
\begin{equation}\label{eq: B1}
C_{10}=\left(1-x_e\right)n_\mathrm{H}\kappa_{1-0}^\mathrm{HH}+x_e\left(n_\mathrm{H}+n_\mathrm{He}\right)\kappa_{1-0}^\mathrm{eH}+x_en_\mathrm{H}\kappa_{1-0}^\mathrm{pH}.
\end{equation}
The $\kappa_{1-0}^\mathrm{iH}$ parameters are the collision rates (in units of $\mathrm{cm^3/sec}$) of hydrogen atoms with particles of species $i$ and are functions of $T_k$ only. In {\tt 21cmFAST}, their tabulated values are taken from~\cite{2005ApJ...622.1356Z, Furlanetto:2006su, Furlanetto:2007te}. Thus, the fractional fluctuation in $x_\mathrm{coll}$ to linear order is
\begin{flalign}\label{eq: B2}
\nonumber&\frac{\delta x_\mathrm{coll}}{\bar  x_\mathrm{coll}}=\delta_b+\bar C_{10}^{-1}\bar n_\mathrm{H}\Bigg\{\left[f^{-1}_\mathrm{H}\bar\kappa^\mathrm{eH}_{1-0}+\bar\kappa^\mathrm{pH}_{1-0}-\bar\kappa^\mathrm{HH}_{1-0}\right]\delta x_e&
\\&\hspace{2mm}+\left[\left(1-\bar x_e\right)\frac{\partial \bar\kappa^\mathrm{HH}_{1-0}}{\partial \bar T_k}+f^{-1}_\mathrm{H}\bar x_e\frac{\partial \bar\kappa^\mathrm{eH}_{1-0}}{\partial \bar T_k}+\bar x_e\frac{\partial \bar\kappa^\mathrm{pH}_{1-0}}{\partial \bar T_k}\right]\delta T_k\Bigg\},&
\end{flalign}
where $f_\mathrm{H}\equiv\bar n_\mathrm{H}/\left(\bar n_\mathrm{H}+\bar n_\mathrm{He}\right)\approx0.925$. Taking temperature fluctuations into account modifies Eq.~\eqref{eq: 31},
\begin{eqnarray}\label{eq: B3}
\nonumber\frac{\delta T_s}{\bar T_s}&=&-\frac{\left(1-\bar T_k/T_\gamma\right)\bar x_\mathrm{coll}\bar x_\mathrm{CMB}}{\left(\bar x_\mathrm{CMB}+\bar x_\mathrm{coll}\right)\left(\bar x_\mathrm{CMB}\bar T_k/T_\gamma +\bar x_\mathrm{coll}\right)}\frac{\delta x_\mathrm{coll}}{\bar  x_\mathrm{coll}}
\\&&+\frac{\bar x_\mathrm{coll}}{\bar x_\mathrm{CMB}\bar T_k/T_\gamma+\bar x_\mathrm{coll}}\frac{\delta T_k}{\bar T_k},
\end{eqnarray}
while taking ionization fluctuations into account modifies Eq.~\eqref{eq: 33},
\begin{equation}\label{eq: B4}
\delta_{21}=\delta_b-\frac{\delta T_s}{\bar T_s}\frac{1}{1-\bar T_s/T_\gamma}-\frac{1}{H}\frac{d\left(\mathbf{\hat n}\cdot\mathbf{v}_b\right)}{d\left(\mathbf{\hat n}\cdot\mathbf{x}\right)}+\frac{\delta x_e}{1-\bar x_e}.
\end{equation}
Then, combining Eqs.~\eqref{eq: B2}-\eqref{eq: B4} and using the definitions of $c_T$, $c_{x_e}$ and $c_\mathrm{21,iso}$ results in Eq.~\eqref{eq: 41}. We note that Eq.~\eqref{eq: B3} is consistent with Eq.~(32) in LC07 when the small $\mathcal O\left(\tau_{21}\right)$ term and CMB temperature fluctuations are ignored.

To account for the fluctuations in the optical depth, we need to modify Eq.~\eqref{eq: B3}, as fluctuations in $x_\mathrm{CMB}$ contribute to the fluctuations in $T_s$,
\begin{flalign}\label{eq: B5}
\nonumber&\frac{\delta T_s}{\bar T_s}\to\frac{\delta T_s}{\bar T_s}&
\\&\hspace{5mm}+\frac{\left(1-\bar T_k/T_\gamma\right)\bar x_\mathrm{coll}\bar x_\mathrm{CMB}}{\left(\bar x_\mathrm{CMB}+\bar x_\mathrm{coll}\right)\left(\bar x_\mathrm{CMB}\bar T_k/T_\gamma+\bar x_\mathrm{coll}\right)}\frac{\delta x_\mathrm{CMB}}{\bar x_\mathrm{CMB}},&
\end{flalign}
where to linear order,
\begin{equation}\label{eq: B6}
\frac{\delta x_\mathrm{CMB}}{\bar x_\mathrm{CMB}}=\delta\left[\ln\left(\frac{1-\mathrm{e}^{-\tau_{21}}}{\tau_{21}}\right)\right]=\left(\frac{\bar\tau_{21}}{\mathrm{e}^{\bar\tau_{21}}-1}-1\right)\frac{\delta\tau_{21}}{\bar \tau_{21}}.
\end{equation}
The fractional fluctuations in the optical depth, $\delta\tau_{21}/\bar\tau_{21}$, are given by an expression very similar to Eq.~\eqref{eq: B4}, but not exactly the same,
\begin{equation}\label{eq: B7}
\frac{\delta\tau_{21}}{\bar \tau_{21}}=\delta_b-\frac{\delta T_s}{\bar T_s}-\frac{1}{H}\frac{d\left(\mathbf{\hat n}\cdot\mathbf{v}_b\right)}{d\left(\mathbf{\hat n}\cdot\mathbf{x}\right)}+\frac{\delta x_e}{1-\bar x_e}.
\end{equation}
In addition, the assumption $\tau_{21}\ll 1$ in Eq.~\eqref{eq: 32} has to be relaxed,
\begin{eqnarray}\label{eq: B8}
\nonumber T_{21}&\propto&\left(1-x_e\right)n_\mathrm{H}\left(1-\frac{T_\gamma}{T_s}\right)\left(1+\frac{1}{H}\frac{d\left(\mathbf{\hat n}\cdot\mathbf{v}_b\right)}{d\left(\mathbf{\hat n}\cdot\mathbf{x}\right)}\right)^{-1}
\\&&\times\frac{1-\mathrm{e}^{-\tau_{21}}}{\tau_{21}},
\end{eqnarray}
where again the factor of proportionality contains terms that are uniform in space. As a consequence, Eq.~\eqref{eq: B4} is modified,
\begin{eqnarray}\label{eq: B9}
\nonumber\delta_{21}&=&\delta_b-\frac{\delta T_s}{\bar T_s}\frac{1}{1-\bar T_s/T_\gamma}-\frac{1}{H}\frac{d\left(\mathbf{\hat n}\cdot\mathbf{v}_b\right)}{d\left(\mathbf{\hat n}\cdot\mathbf{x}\right)}+\frac{\delta x_e}{1-\bar x_e}
\\&&+\delta\left[\ln\left(\frac{1-\mathrm{e}^{-\tau_{21}}}{\tau_{21}}\right)\right].
\end{eqnarray}
It should be noted that Eqs.~\eqref{eq: B6}-\eqref{eq: B9} are consistent with Eq.~(22) in LC07.

Combining Eqs.~\eqref{eq: B6} and \eqref{eq: B9} yields the following modification to $\delta_{21}$,
\begin{equation}\label{eq: B10}
\delta_{21}\to\delta_{21}+\left(\frac{\bar\tau_{21}}{\mathrm{e}^{\bar\tau_{21}}-1}-1\right)\frac{\delta\tau_{21}}{\bar \tau_{21}}.
\end{equation}
Because of the peculiar velocity term  in Eq.~\eqref{eq: B7}, we need to modify our definition of $c_\mathrm{21,iso}$ in Eq.~\eqref{eq: 34},
\begin{equation}\label{eq: B11}
\delta_{21}\equiv c_\mathrm{21,iso}\delta_b-\frac{\bar\tau_{21}}{\mathrm{e}^{\bar\tau_{21}}-1}\frac{1}{H}\frac{d\left(\mathbf{\hat n}\cdot\mathbf{v}_b\right)}{d\left(\mathbf{\hat n}\cdot\mathbf{x}\right)}.
\end{equation}
It can be shown that $\delta\tau_{21}/\left(\bar\tau_{21}\cdot\delta_b\right)$ is positive (i.e.\ overdense regions lead to a stronger optical depth), and since the expression in the parentheses of Eq.~\eqref{eq: B10} can be Taylor expanded to $-\bar\tau_{21}/2$, we conclude that $c_\mathrm{21,iso}$ receives a negative contribution from the optical depth (see Fig.~\ref{Fig: figure_8}). Also, we note that the factor $\bar\tau_{21}/\left(\mathrm{e}^{\bar\tau_{21}}-1\right)\sim1-\bar\tau_{21}/2$ has to be properly inserted into Eqs.~\eqref{eq: 38} and \eqref{eq: 39} when the 21-cm power spectrum is evaluated.

\vspace{-0.15in}

\section{Disentangling $T_s$ and $x_\mathrm{CMB}$}\label{sec: Disentangling Ts and xCMB}
\vspace{-0.05in}

An inevitable challenge arises when attempting to calculate $T_s$ in Eq.~\eqref{eq: 30}. It depends on $x_\mathrm{CMB}$, while $x_\mathrm{CMB}$ depends on $\tau_{21}$ which depends on $T_s$. Therefore, $T_s$ and $x_\mathrm{CMB}$ are mathematically entangled.

To overcome this difficulty numerically, we use the value of $T_s$ from the previous iteration in calculating the current value of the optical depth. This is a very good approximation provided that the redshift step is small enough and $T_s$ remains roughly the same between two consecutive redshift iterations (see another approach of solution by iterations in Refs.~\cite{Reis:2021nqf, Fialkov:2019vnb}). As for the initial value of $T_s$, we expand $x_\mathrm{CMB}$ to first order in $\tau_{21}$,
\begin{eqnarray}\label{eq: C1}
\nonumber T_s&\approx&\frac{1-\frac{\tau_{21}}{2}+x_\mathrm{coll}}{\left(1-\frac{\tau_{21}}{2}\right)T_{\gamma}^{-1}+T_{k}^{-1}x_\mathrm{coll}}
\approx\frac{1+x_\mathrm{coll}}{T_{\gamma}^{-1}+T_{k}^{-1}x_\mathrm{coll}}
\\&&-\frac{\tau_{21}}{2}\frac{\left(1-T_k/T_{\gamma}\right)x_\mathrm{coll}}{\left(T_\gamma^{-1}+T_k^{-1}x_\mathrm{coll}\right)\left(T_k/T_\gamma+x_\mathrm{coll}\right)}.
\end{eqnarray}
Using the zeroth order in $\tau_{21}$, the solution to $T_s$ can be written in the following form
\begin{equation}\label{eq: C2}
1-\frac{T_s}{T_\gamma}=\frac{\left(1-T_k/T_\gamma\right)x_\mathrm{coll}}{T_k/T_\gamma+x_\mathrm{coll}}.
\end{equation}
Combining Eqs.~\eqref{eq: C1}-\eqref{eq: C2}, it follows that
\begin{eqnarray}\label{eq: C3}
\nonumber T_s&=&\frac{1+x_{\mathrm{coll}}}{T_{\gamma}^{-1}+T_{k}^{-1}x_{\mathrm{coll}}}
\\&&\hspace{-5mm}-\frac{1}{2}\tau_{21}T_{s}\left(\frac{T_{s}^{-1}}{T_{\gamma}^{-1}+T_{k}^{-1}x_{\mathrm{coll}}}-\frac{T_k/T_\gamma}{T_k/T_\gamma+x_{\mathrm{coll}}}\right).~~~~~~
\end{eqnarray}
Now, using once again the zeroth order solution to $T_s$ in the RHS of Eq.~\eqref{eq: C3}, we arrive at
\begin{eqnarray}\label{eq: C4}
\nonumber T_s=\frac{1+x_{\mathrm{coll}}}{T_{\gamma}^{-1}+T_{k}^{-1}x_{\mathrm{coll}}}
-\frac{1}{2}\frac{\tau_{21}T_{s}\left(1-T_k/T_\gamma\right)x_\mathrm{coll}}{\left(T_k/T_\gamma+x_\mathrm{coll}\right)\left(1+x_\mathrm{coll}\right)}.\\
\end{eqnarray}
It is worthwhile to note that Eq.~\eqref{eq: C4} is identical to Eq.~(31) in LC07, thus their expression for $T_s$ is the first order solution in $\tau_{21}$. Because $\tau_{21}\propto T_s^{-1}$, Eq.~\eqref{eq: C4} is a closed form solution to $T_s$, and we use it to set the initial value for $T_s$ (which at recombination is $\approx T_\gamma$).

The mathematical entanglement of $T_s$ and $\tau_{21}$ also poses another challenge when calculating their fluctuations analytically. From Eqs.~\eqref{eq: B5}-\eqref{eq: B6}, $\delta T_s$ depends on $\delta\tau_{21}$, but from Eq.~\eqref{eq: B7} the latter depends on the former. Since we have two equations with two unknowns, the disentanglement is straightforward, although tedious:
\begin{widetext}
\begin{eqnarray}\label{eq: C5}
\nonumber\left(\frac{1}{1-\bar{T}_{s}/T_{\gamma}}+\frac{\bar{x}_{\mathrm{CMB}}F\left(\bar{\tau}_{21}\right)}{\bar{x}_{\mathrm{CMB}}+\bar{x}_{\mathrm{coll}}}\right)\frac{\delta T_{s}}{\bar{T}_{s}}&=&-\frac{\bar{x}_{\mathrm{CMB}}}{\bar{x}_{\mathrm{CMB}}+\bar{x}_{\mathrm{coll}}}\frac{\delta x_{\mathrm{coll}}}{\bar{x}_{\mathrm{coll}}}+\frac{1}{1-\bar{T}_{k}/T_{\gamma}}\frac{\delta T_{k}}{\bar{T}_{k}}
\\&&+\frac{\bar{x}_{\mathrm{CMB}}F\left(\bar{\tau}_{21}\right)}{\bar{x}_{\mathrm{CMB}}+\bar{x}_{\mathrm{coll}}}\left(\delta_{b}-\frac{1}{H}\frac{d\left(\mathbf{\hat{n}}\cdot\mathbf{v}_{b}\right)}{d\left(\mathbf{\hat{n}}\cdot\mathbf{x}\right)}+\frac{\delta x_{e}}{1-\bar{x}_{e}}\right),
\end{eqnarray}
\end{widetext}
where we have defined for brevity
\begin{equation}\label{eq: C6}
F\left(\bar{\tau}_{21}\right)\equiv\frac{\bar{\tau}_{21}}{e^{\bar{\tau}_{21}}-1}-1=-\frac{\bar\tau_{21}}{2}+\mathcal O\left(\bar\tau_{21}^2\right).
\end{equation}
As a consistency check, note that Eq.~\eqref{eq: C5} reduces to Eq.~\eqref{eq: B3} in the limit $\bar\tau_{21}\to0$. The above solution for $\delta T_s$ can now be substituted in the expression for $\delta\tau_{21}$, Eq.~\eqref{eq: B7}. In addition, because $\delta T_s$ explicitly depends on the peculiar velocity, our definition to $c_\mathrm{21,iso}$ has to be further modified,
\begin{flalign}\label{eq: C7}
\nonumber&\delta_{21}\equiv c_\mathrm{21,iso}\delta_b-\frac{1}{H}\frac{d\left(\mathbf{\hat n}\cdot\mathbf{v}_b\right)}{d\left(\mathbf{\hat n}\cdot\mathbf{x}\right)}\Bigg[1+F\left(\bar\tau_{21}\right)&
\\&\hspace{2mm}+\frac{\bar{x}_{\mathrm{CMB}}F\left(\bar{\tau}_{21}\right)}{\bar{x}_{\mathrm{CMB}}+\bar{x}_{\mathrm{coll}}+\left(1-\bar{T}_{s}/T_{\gamma}\right)\bar{x}_{\mathrm{CMB}}F\left(\bar{\tau}_{21}\right)}\Bigg],&
\end{flalign}
and we stress that the expression in the brackets has to be accounted for when evaluating Eqs.~\eqref{eq: 38} and \eqref{eq: 39}.

\bibliography{21cmFirstCLASS_II_early_linear_fluctuations_21cm_signal.bib}

\end{document}